\documentclass[a4paper, amsfonts, amssymb, amsmath, reprint, showkeys, nofootinbib,longbibliography]{revtex4-1}
\usepackage[english]{babel}
\usepackage[utf8]{inputenc}
\usepackage{amsthm}
\usepackage{mathtools}
\usepackage{physics}
\usepackage{xcolor}
\usepackage{graphicx}
\usepackage{siunitx}
\usepackage[left=23mm,right=13mm,top=35mm,columnsep=15pt]{geometry} 
\usepackage{adjustbox}
\usepackage{placeins}
\usepackage[T1]{fontenc}
\usepackage{lipsum}
\usepackage{csquotes}
\usepackage[compat=1.0.0]{tikz-feynman}
\usepackage{feynmp-auto}
\usepackage[version=4]{mhchem}
\usepackage{prettyref}
\usepackage[pdftex, pdftitle={Article}, pdfauthor={Author}]{hyperref} 

\newcommand{\ii}{\mathrm{i}}
\newcommand{\ee}{\mathrm{e}}

\newcommand*{\citesec}[1]{\S~{#1}}
\newcommand*{\citechap}[1]{Ch.~{#1}}
\newcommand*{\citefig}[1]{Fig.~{#1}}
\newcommand*{\citetable}[1]{Table~{#1}}

\newcommand*{\citefootnote}[1]{fn.~{#1}}

\newrefformat{sec}{\citesec{\ref{#1}}}
\newrefformat{fig}{\citefig{\ref{#1}}}
\newrefformat{tbl}{\citetable{\ref{#1}}}
\newrefformat{chap}{\citechap{\ref{#1}}}
\newrefformat{fn}{\citefootnote{\ref{#1}}}
\newrefformat{box}{Box~\ref{#1}}
\newrefformat{ex}{\ref{#1}}

\begin{document}

\title{
Data-driven Low-rank Approximation for Electron-hole Kernel and Acceleration of Time-dependent GW Calculations
}
\author{Bowen Hou{$^{1,\dagger}$}}
\author{Jinyuan Wu{$^{1,\dagger}$}}
\author{Victor Chang Lee{$^1$}}
\author{Jiaxuan Guo{$^1$}}
\author{Luna Y. Liu{$^2$}}
\author{Diana Y. Qiu$^{1,}$}
\email{diana.qiu@yale.edu}
\affiliation{$^1$ Department of Materials Science, Yale University, New Haven, CT, 06511, USA}
\affiliation{$^2$ Department of Applied Physics, Yale University, New Haven, CT, 06511, USA}

\def\thefootnote{$^\dagger$}\footnotetext{These authors contributed equally to this work}
\date{\today} 

\begin{abstract}
Many-body electron-hole interactions are essential for understanding non-linear optical processes and ultrafast spectroscopy of materials. Recent first principles approaches based on nonequilibrium Green's function formalisms, such as the time-dependent adiabatic GW (TD-aGW) approach, can predict the nonequilibrium dynamics of excited states including electron-hole interactions.
However, the high dimensionality of the electron-hole kernel poses significant computational challenges for scalability. 
Here, we develop a data-driven low-rank approximation for the electron-hole kernel, leveraging localized excitonic effects in the Hilbert space of crystalline systems. Through singular value decomposition (SVD) analysis, we show that the subspace of non-zero singular values, containing the key information of the electron-hole kernel, retains a small size even as the k-grid grows, ensuring computational feasibility with extremely dense k-grids for converged calculations. 
Utilizing this low-rank property, we achieve at least 95\% compression of the kernel and an order-of-magnitude speedup of TD-aGW calculations.
Our method, rooted in physical interpretability, outperforms existing machine learning approaches by avoiding intensive training processes and eliminating time-accumulated errors, providing a general framework for high-throughput, nonequilibrium simulation of light-driven dynamics in materials.

\end{abstract}

\maketitle

\section{Introduction}

In recent years, ultrafast spectroscopic techniques, including pump-probe transient absorption \cite{leone2014will,ramasesha2016real}, solid state high-harmonic generation (HHG) \cite{liu2017high}, terahertz spectroscopy \cite{ulbricht2011carrier}, and time-resolved angle-resolved photoemission spectroscopy (TR-ARPES) \cite{boschini2024time}, have emerged as crucial techniques for studying novel non-equilibrium physics in condensed matter systems.
Interpreting the complex features observed in these experiments requires a predictive understanding of the roles of electron-electron and electron-phonon interactions and other many-body effects under highly non-equilibrium and strong-field conditions.
Electron-electron (and electron-hole) interactions, in particular, play a predominant role in the coherent femtosecond dynamics of solid state systems, leading to a wide variety of phenomena, including light-induced superconductivity, many-body enhancement of HHG, coherent and incoherent excitonic effects in ARPES, and exciton-enhanced shift currents \cite{cavalleri2018photo,liu2017high,chang2024many,chan2021giant,chan2023giant,man2021experimental,madeo2020directly,lin2022exciton}.
In the last decades, there has been an intense effort to simulate these driven non-equilibrium electronic processes from first principles \cite{attaccalite2011real,perfetto2015nonequilibrium,perfetto2022real,marques2004time,reining2002excitonic,romaniello2009double}.
One highly successful approach is the time-dependent adiabatic $GW$ (TD-aGW) \cite{attaccalite2011real,sangalli2021excitons,chan2021giant,chan2023giant}, which is a non-equilibrium quantum master equation approach that simulates the dynamics of the single-electron reduced density matrix and has successfully described diverse phenomena, including transient absorption \cite{sangalli2021excitons},  TR-ARPES~\cite{chan2023giant}, and HHG with exciton effects \cite{chang2024many}. 
However, a significant challenge lies in the computational demands of these simulations, which are prohibitively slow for large systems and for parameter sweeping, which is necessary to describe different driving-field conditions in ultrafast and strong-field spectroscopic measurements.

There has been growing interest in leveraging data-driven and machine learning techniques to accelerate first-principles many-body simulations, using methods such as low-rank approximations\cite{luo2024data, lu2015compression, shao2016low, Kaye2022GreenFunction, dunlap2000robust, pham2019compressing,del2019static}, representation learning\cite{zang2024machine, Shinaoka2017, zadoks2024spectral, knosgaard2022representing, hou2024unsupervised, luchnikov2019variational}, equivariant graph neural networks\cite{batzner20223, thomas2018tensor,li2024deep} and other deep learning approaches\cite{yang2020deep, medvidovic2024neural, noe2019boltzmann, carrasquilla2021use}. 
For non-equilibrium simulations, data-driven prediction based on extrapolation from an existing time sequence, using methods such as dynamic mode decomposition (DMD)\cite{yin2023analyzing,maliyov2024dynamic,reeves2023dynamic} or Gaussian processes\cite{gu2024probabilistic}, are commonly employed. However, since these approaches are extrapolative, they fail to describe situations where the external field changes dramatically outside the initially observed time interval, which is crucial for modeling ultrafast spectroscopy. Operator learning neural networks\cite{lu2021learning,li2020fourier, kovachki2023neural,li2020neural,li2022transformer,zhu2024predicting, reeves2024performance} show promise for solving differential equations under more general conditions. 
However, such operator learning requires extensive training datasets for different driving conditions and still suffers from time-accumulated predictive errors and a black-box process that can make it difficult to extract physical understanding\cite{brunton2022data, lu2021learning, shwartz2017opening}.
Thus, the development of new data-driven techniques that can accelerate calculations of many-body dynamics under highly non-equilibrium conditions and is robust against both rapid changes in the external field and error accumulation over the simulation time is urgently needed.

Here, we develop a data-driven low-rank approximation for the electron-hole kernel---the key interaction that encodes exciton effects across different many-body approaches---based on singular value decomposition (SVD). It exploits the inherent localization of excitonic effects within valleys in reciprocal space, which gives rise to 
a low-rank structure for the electron-hole interaction kernel across crystalline materials. In practice, we find that this kernel can be compressed by as much as 95\% in most crystalline solids, while retaining information essential for reproducing both optical absorption spectra, for excitons with near-zero momentum, and energy loss spectra for finite momentum excitons. 
Moreover, we apply the SVD-compressed electron-hole kernel to accelerate TD-aGW calculations,
achieving an order of magnitude speedup for calculations of both linear and nonlinear spectra under different driving conditions. This speedup makes repeated calculations under different field conditions computationally tractable, allowing first principles simulations to mirror experimental conditions.
In contrast with approaches based on extrapolation from early-time dynamics, our physically interpretable approach exhibits no accumulated predictive error over time and requires no datasets or training,
laying the foundation for achieving high throughput simulations of nonequilibrium light-driven dynamics in materials.

\section{Results}\label{band}

\subsection{Two-particle electron-hole kernel in non-equilibrium dynamics}
\label{sec:tdagw}

In interacting many-body systems, the equation of motion of the non-equilibrium single-particle Green's function on the Keldysh contour takes the form:

\begin{equation} 
    \left( \ii \dv{t} -H(t) \right) G(t,t') = \delta(t,t')+\int_C\Sigma(t, \bar{t})G(\bar{t},t') \dd\bar{t},
    \label{eq:KBEgeneral}
\end{equation}
where $G(t, t')$ is a $2\times 2$ matrix containing the retarded, advanced, lesser, and greater two-time Green's functions; $\Sigma$ is the self-energy matrix in the Keldysh formalism
\cite{kremp2005quantum}; and $H(t)$ is a mean-field Hamiltonian.
There is an equivalent adjoint equation for the time-evolution over $t'$, and the set of self-consistent equations derived from the Eq.~\ref{eq:KBEgeneral} is referred to as the Kadanoff-Baym equations (KBE) \cite{kadanoff2018quantum,kremp2005quantum,kadanoff1962green}. The self-energy can be computed within conserving schemes, such as the $GW$ approximation \cite{hedin1965new,hedin1970effects,hybertsen1986electron}—i.e. $\Sigma=\ii GW$, where $W$ is the screened Coulomb potential calculated in the random phase approximation (RPA)\cite{adler1962quantum,wiser1963dielectric,ceperley1980ground}. 

One challenge in solving Eq.~\ref{eq:KBEgeneral} is that the time evolution of $G$ over $t$ and $t'$ must be performed simultaneously. 
In first principles calculations, a common approximation is to decouple the many-body effects in equilibrium from the changes induced by the driving field by splitting the self-energy into an equilibrium piece $\Sigma[G_0]$, plus a correction term $\delta\Sigma[G, G_0]=\Sigma[G]-\Sigma[G_0]$, where $G_0$ is the Green's function in the absence of an external field. Then, $G_0$ is evaluated within the equilibrium $GW$ approximation \cite{hedin1965new,hedin1970effects,hybertsen1986electron}, while the correction is evaluated in the static limit of the $GW$ approximation (i.e., the static Coulomb hole and screened exchange approximation, or static-COHSEX) which ignores retardation effects in $\Sigma^{GW}$ and thus eliminates the double time dependence \cite{hedin1965new,hedin1970effects,hybertsen1986electron,attaccalite2011real,sangalli2021excitons}.
In this approximation, the non-equilibrium dynamics of the system can be described by the equation of motion of the reduced single-particle density matrix, expressed here in a Bloch basis, $\rho_{nm\textbf{k}} = \bra{n\textbf{k}}\rho\ket{m\textbf{k}}$ \cite{attaccalite2011real}:
\begin{widetext}
\begin{equation}
\begin{aligned}
\ii\frac{\partial}{\partial t}\rho(t)_{nm\textbf{k}} &= \left[ H^\text{QP}  + \delta\Sigma^{GW}(t) - e\textbf{E}(t)\cdot\textbf{r}, \rho(t) \right]_{nm,\textbf{k}} \\
&= (\epsilon^\text{QP}_{n\textbf{k}}-\epsilon^\text{QP}_{m\textbf{k}})\rho(t)_{nm,\textbf{k}} + [\delta\Sigma^{GW}(t), \rho(t)]_{nm\vb{k}} - e \vb{E}(t) [\textbf{r}, \rho(t)]_{nm\vb{k}}, \\
\end{aligned}
\label{eq:tdagw}
\end{equation}
\end{widetext}
where the quasiparticle Hamiltonian $H^\text{QP} = H_{0} + \Sigma^{GW}[G_0]$, and
$\var{\Sigma}^{GW}$ is the non-equilibrium change in the static COHSEX self-energy,
in which time-dependent changes to the screened Coulomb interaction are typically neglected \cite{attaccalite2011real}. 
$\vb{E}(t)$ is an arbitrary external electric field, which couples to the system in the length gauge through the position operator $\vb{r}$ \cite{aversa1995nonlinear,virk2007semiconductor}.
The dipole approximation is made, so there are no finite momentum transitions,
and the reduced single-electron density matrix only needs one crystal momentum index $\vb{k}$.
We refer to Eq.~\eqref{eq:tdagw} as the time-dependent adiabatic GW (TD-aGW)\cite{attaccalite2011real,chan2021giant,chan2023giant}.

\begin{figure}
    \centering
    \includegraphics[width=\linewidth]{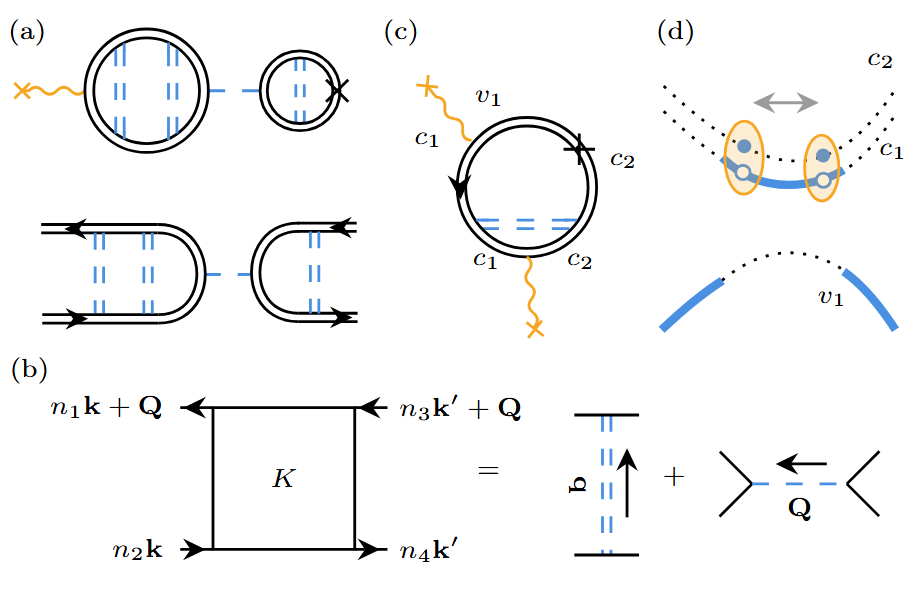}
    \caption{Feynman diagram analysis of the relation between TD-aGW and BSE. Double solid black lines are electron or hole quasiparticle propogators. Double blue dashed lines correspond to RPA-screened Coulomb interactions, and single blue dashed lines correspond to the bare Coulomb interaction. A cross means that the two ends connected to the cross are at the same time, corresponding to the calculation of physical observables like the dipole moment. Orange wavy lines correspond to the external driving field. (a) Linear response of TD-aGW with (top) and without (bottom) the coupling to the external field. Note that the linear response of TD-aGW is a ladder diagram approximation which is equivalent to GW-BSE. 
    (b) $K^{\vb{Q}}_{n_1 n_2 \vb{k} n_3 n_4 \vb{k}'}$ is the irreducible BSE kernel in the electron-hole pair basis. It is composed of a screened direct term and an unscreened exchange term that correspond exactly to the ladder diagrams in (a).
    (c) Example of a nonlinear response diagram of TD-aGW involving $K_{c_1c_2\vb{k}c_1c_2\vb{k}'}$, which will never appear in the equilibrium BSE.
    (d) Schematic of the physical process in (c): the external driving first causes occupation of the conduction bands (corresponding to the blue highlighted part of the conduction band), and then an electron-hole pair between two conduction bands form and coherently transitions to another electron-hole pair due to renormalization by $K_{c_1c_2\vb{k}c_1c_2\vb{k}'}$.
    The polarization of the latter electron-hole pair is then given by (c).}
    \label{fig:tdagw-diagram}
\end{figure}

In the linear response limit, TD-aGW can be explicitly proven to be equivalent to the $GW$-Bethe-Salpeter equation ($GW$-BSE) approach in many-body perturbation theory (\prettyref{fig:tdagw-diagram} (a,b)) \cite{attaccalite2011real,rocca2012solution}.
In $GW$-BSE, exciton eigenstates are found by diagonalizing an effective Hamiltonian consisting of the non-interacting electron-hole pair Hamiltonian plus the BSE electron-hole interaction kernel, which is derived by taking the functional derivative $K = \fdv{\Sigma}{G}$ of the $GW$ self-energy and then ignoring both the non-equilibrium time evolution of the screened Coulomb interaction $W$ and the frequency dependence of the screened Coulomb interaction \cite{rohlfing2000electron,rocca2012solution,albrecht1998ab}.
We note that these approximations are consistent with the approximations made to the $GW$ self-energy in TD-aGW. 
Therefore, under the static COHSEX approximation for the single-electron self-energy in TD-aGW,
$\var{\Sigma}^{GW}$ is given by the matrix product of the BSE kernel $K$ and the non-equilibrium lesser Green's function, i.e. 
\begin{equation}
    \var{\Sigma}_{m n \vb{k}}^{GW} = \sum_{m', n', \vb{k}'} K_{m n \vb{k} m' n' \vb{k}'} \rho_{m' n' \vb{k}'} .
    \label{eq:sigma-from-k}
\end{equation}
The non-equilibrium electron dynamics are therefore completely controlled by the two-particle BSE kernel $K_{n_1 n_2 \vb{k} n_3 n_4 \vb{k}'}$ (\prettyref{fig:tdagw-diagram}). 
When finite momentum electron-hole pairs are created,
$K_{n_1 n_2 \vb{k} n_3 n_4 \vb{k}'}$ has to be replaced by $K_{n_1 n_2 \vb{k}, n_3 n_4 \vb{k}'}^{ \vb{Q}}$,
where $\vb{k}$ and $\vb{k}'$ are the momenta of the hole after and before electron-hole interaction,
and $\vb{Q}$ is the center of mass momentum of the electron-hole pair.
The kernel is composed of two terms (\prettyref{fig:tdagw-diagram}(b)) \cite{BGW1, Rohlfing2000, strinati1988application}: an exchange interaction mediated by the bare Coulomb interaction $v$:
\begin{equation}
    \begin{aligned}
        &K^{\text{x}, \vb{Q}}_{n_1 n_2 \vb{k} n_3 n_4 \vb{k}'} \coloneqq \langle n_1n_2 \vb{k} \vb{Q} | K^\text{x} | n_3 n_4 \vb{k}' \vb{Q} \rangle  \\
        =& \sum_{\vb{G}} M_{n_1n_2}(\vb{k}, \vb{Q}, \vb{G}) v(\vb{Q} + \vb{G}) M^*_{n_3 n_4}(\vb{k}', \vb{Q}, \vb{G}), 
    \end{aligned}
    \label{eq:bse-exchange-def}
\end{equation}
and a direct interaction mediated by the screened Coulomb interaction
\begin{equation}
    \begin{aligned}
        &K^{\text{d}, \vb{Q}}_{n_1 n_2 \vb{k} n_3 n_4 \vb{k}'} \coloneqq \langle n_1n_2 \vb{k} \vb{Q} | K^\text{d} | n_3' n_4' \vb{k}' \vb{Q} \rangle  \\
        =& - \sum_{\vb{G} \vb{G}'} M_{n_1n_3}(\vb{k}' + \vb{Q}, \vb{q}, \vb{G}) W_{\vb{G} \vb{G}'}(\vb{q}) M_{n_2 n_4}^*(\vb{k}', \vb{q}, \vb{G}').\\
    \end{aligned}
    \label{eq:bse-direct-def}
\end{equation}
Here, the matrices are written in the electron-hole pair basis, where $\ket{nn' \vb{k} \vb{Q}}=\ket{n \vb{k} + \vb{Q}}_{\text{electron}} \otimes \ket{n' \vb{k}}_{\text{hole}}$; $v(\vb{Q}+\vb{G})$ denotes the bare Coulomb interaction, where $\vb{G}$ is a reciprocal lattice vector representing how much the momentum transfer exceeds the first Brillouin zone; $W_{\vb{G} \vb{G}'}(\vb{q})$ is the screened Coulomb interaction; $\vb{q}=\vb{k}-\vb{k}'$; and $M_{nn'}(\vb{k}, \vb{q}, \vb{G}) \coloneqq \langle n \vb{k} + \vb{q} | \ee^{\ii(\vb{q} + \vb{G}) \cdot \vb{r}} | n' \vb{k} \rangle$.
In the current work (and previous first principles implementations \cite{Giannozzi_2009,BGW1}), the electron and hole states are Kohn-Sham (KS) states \cite{kohn1965self} derived from density functional theory (DFT) in a plane-wave basis.

We note that the equilibrium BSE is a ladder diagram approximation and can be viewed as a self-energy correction to the free electron-hole pair, and therefore only matrix elements in the form of $K_{cv c' v'}$, $K_{c v v' c'}$, $K_{v c v' c'}$ and $K_{v c c' v'}$ appear (indexing convention in \prettyref{fig:tdagw-diagram}(b)),
corresponding to interactive renormalization of the resonant and anti-resonant parts of the free electron-hole pair propagator,
where $c$ and $v$ run over the conduction subspace and the valence subspace of the system, respectively
\cite{albrecht1998ab,rohlfing2000electron}.
Nonlinear TD-aGW, on the other hand, does not have such constraints, and the whole $K^{\vb{Q}}_{n_1 n_2 \vb{k}, n_3 n_4 \vb{k}'}$ matrix contributes to the time evolution (\prettyref{fig:tdagw-diagram}(c)).
The intuition is that in nonlinear response, the external driving field can first pump electrons to conduction bands and then create electron-hole pairs from the occupied conduction band (\prettyref{fig:tdagw-diagram}(d)).
In non-equilibrium BSE, which can be seen as the incoherent limit of TD-aGW in a pump-probe scheme, these components also appear \cite{perfetto2015nonequilibrium}.

\subsection{SVD analysis of electron-hole kernel}\label{sec:bse-kernel-svd}

The electron-hole kernel matrix has a size of $N_{\text{b}}^4 N_{\vb{k}}^2 N_{\vb{Q}}$, where $N_{\text{b}}$, $N_{\vb{k}}$ and $N_{\vb{Q}}$ denote the number of bands, the number of $\vb{k}$-points, and the number of momenta $\vb{Q}$ of the electron-hole pairs, respectively. For a typical solid, $N_{\vb{k}}$ is on the order of $10^3\text{-}10^6$ and the number of bands $N_{\text{b}}$ is $1\text{-}10^2$. Here, we start by only considering coupling to an external field in the dipole approximation, so $N_{\vb{Q}}=1$, but more generally, when scattering to finite momenta is allowed  $N_{\vb{Q}}$ is of the same order as $N_{\vb{k}}$.
The kernel matrix is therefore numerically demanding to evaluate
and also poses computational challenges for downstream tasks. 

In the equilibrium BSE, the exciton envelope wavefunction is often localized in a low-energy region of the Brillouin Zone (i.e. a valley) so that the BSE can be solved on a nonuniform patch in k-space 
\cite{alvertis2023importance, Hou2023Bi2Se3,li2017direct,Wu2024WTe2, Qiu2012MoS2}.
However, in TD-aGW (Eq.~\eqref{eq:tdagw}), the external field can drive intraband transitions for the excited electrons and holes,
and their motion may no longer be restricted to a single clearly-defined region of $\vb{k}$-space, determined by the exciton eigenstate (\prettyref{fig:tdagw-diagram}(d)). 
The occurrence of exciton patches nonetheless hints at a low-rank structure of the BSE kernel (Appendix B), which is explored in this section.

\begin{figure}
    \centering
    \includegraphics[width=\linewidth]{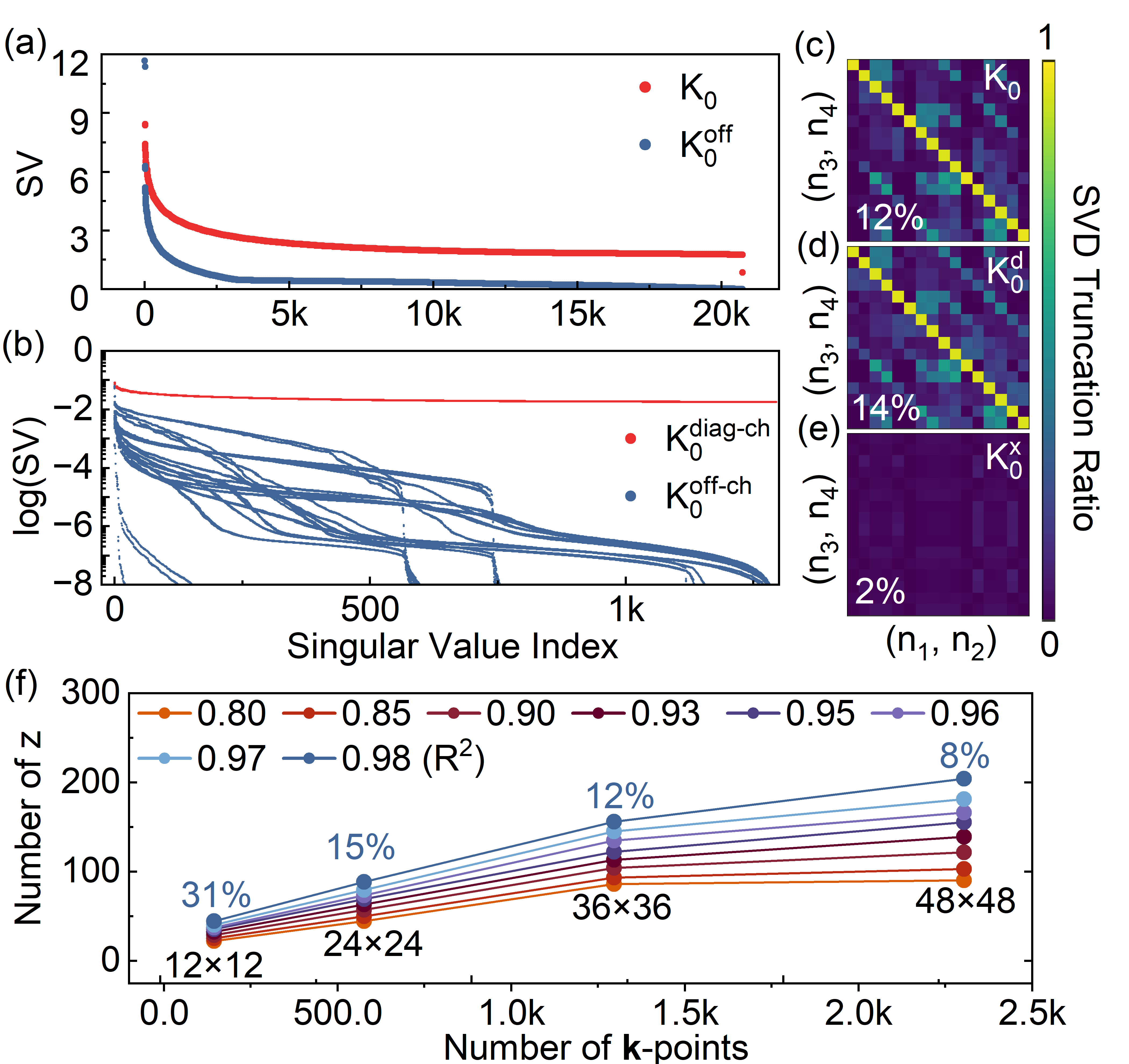}
    \caption{(a) Decay of sorted singular values for the electron-hole kernel ${K_0}$ of monolayer MoS$_2$. The red dots represent the singular values of the full ${K_0}$, and the blue dots represent the singular values of the off-diagonal part ${K^{\text{off}}_{0}}$. (b) Zoomed in logscale decay of the sorted singular values for each channel $K_0^\text{ch}$. The red dots represent the diagonal channels  $K_0^\text{diag-ch}$, and the blue dots represent the off-diagonal channels $K_0^\text{off-ch}$. (c-e) Heat maps of the minimum ratio of singular values required to reconstruct each channel with an accuracy of $R^2=0.98$ for $K_0$, the direct interaction $K_{0}^\text{d}$, and the exchange interaction $K_{0}^\text{x}$. The mean truncation ratios for the SVD of the $K_0^\text{off-ch}$ are 12\%, 14\%, and 2\% respectively. (f) The minimum number of $z$ (preserved singular values) for reconstructing $K_0$ starting from different k-grids of $12\times12\times1$, $24\times24\times1$, $36\times36\times1$, and $48\times48\times1$. The reconstruction accuracy ranges from $R^2=$0.80 to 0.98, represented by different colors. The blue labels 31\%, 15\%, 12\%, and 8\% denote compression rate with reconstruction accuracy of $R^2=$0.98 for different k-grids.}
    \label{fig:svd_analysis}
\end{figure}

We start by focusing on the photo-excited electron-hole kernel $K_0 \coloneqq K^{\vb{Q}=0}_{n_1 n_2 \vb{k}, n_3 n_4 \vb{k}'} \in \mathbb{C}^{N_{\vb{k}}N_{\text{b}}^2 \times N_{\vb{k}}N_{\text{b}}^2}$ of monolayer $\ce{MoS2}$, a prototypical 2D material known for hosting strong exciton effects~\cite{qiu2013optical}. More computational details are given in Appendix A. $K_0$ can be separated into contributions from transitions between different bands, which we refer to as channels $K_0^\text{ch} \in \mathbb{C}^{N_{\vb{k}} \times N_{\vb{k}}}$, where the channel, $\text{ch}$, is defined as $(n_1, n_2) \to (n_3, n_4)$. 
A detailed analysis of individual channels (see Appendix B) reveals that off-diagonal channels $K_0^\text{off-ch}$, involving transitions between different sets of bands (i.e. $n_1, n_2 \neq n_3, n_4$), are low-rank. However, diagonal channels $K_0^\text{diag-ch}$ between the same set of bands (i.e. $n_1, n_2=n_3, n_4$) are close to full-rank because of large diagonal ($\vb{k} = \vb{k}'$) elements. As a result, the entire $K_0$ has a full-rank structure, making it difficult to achieve significant compression when applying a naive SVD directly to $K_0$.

To address this, we further decompose $K_0$ into parts that are diagonal in $\vb{k}$, $K_0^\text{diag}$ (i.e. $n_1, n_2, \vb{k} = n_3, n_4, \vb{k}'$),  and off-diagonal in $\vb{k}$, $K_0^\text{off}$ (i.e. $n_1, n_2, \vb{k} \neq n_3, n_4, \vb{k}'$). $K_0^\text{diag}$ contributes to most of the non-zero singular values but only accounts for a small portion $1/N_\text{b}^2N_{\vb{k}}$ of the entire $K_0$. The primary computational challenge lies in $K_0^\text{off}$, which is low rank and can be effectively compressed by SVD.
The final low-rank approximation is achieved by performing SVD on $K_0^\text{off}$ and retaining only the first $z$ largest singular values so that the kernel is approximated as 
\begin{equation}
    \Tilde{K_0} \approx K^\text{diag}_0 + \Tilde{U}_0\Tilde{M}_0\Tilde{V}_0^T
\label{eq:svd_diag_offdiag}
\end{equation}
where $ \Tilde{M}_0  \in \mathbb{C}^{z \times z}$, $\Tilde{U}_0 \in \mathbb{C}^{N_{\text{b}}^2 N_{\vb{k}} \times z}$ and $\Tilde{V}_0^T \in \mathbb{C}^{z \times N_{\text{b}}^2 N_{\vb{k}}}$ are the truncated singular value and singular vector matrices for $K_0^\text{off}$.  This low-rank approximation can achieve a compression rate of $z/N_{\text{b}}^2 N_{\vb{k}}$, where $z$ is numerically determined from the computational convergence of $K_0$ reconstructed from $\Tilde{K_0}$.
The sorted singular values of both the entire $K_0$ and $K_0^\text{off}$ are shown in Fig.~\ref{fig:svd_analysis}(a).
After an initial rapid decay, the singular values of $K_0$ converge to a non-zero value, preventing accuracy-preserving truncation. The singular values of the $K_0^\text{off}$, on the other hand, rapidly decay to zero, thus supporting our assumption that \eqref{eq:svd_diag_offdiag} is an efficient low-rank approximation of $K_0$.

Next, to quantitatively analyze the rank of each individual channel in $K_0$, we conduct channel-wise SVD on each $K_0^\text{ch}$ over $\vb{k}$ and $\vb{k}'$, then preserve the $z'$ largest singular values:
\begin{equation}
\begin{aligned}
    \Tilde{K_0^\text{ch}} \approx  \Tilde{U}_0^\text{ch} \Tilde{M}_0^\text{ch} (\Tilde{V}_0^\text{ch})^T
\label{eq:svd-channel-wise}
\end{aligned}
\end{equation}
where $\Tilde{M}_0^\text{ch} \in \mathbb{C}^{z' \times z'}$, $\Tilde{U}_0^\text{ch} \in \mathbb{C}^{N_{\vb{k}} \times z'}$ and
$(\Tilde{V}_0^\text{ch})^T \in \mathbb{C}^{z' \times N_{\vb{k}}}$ are the truncated singular value and singular vector matrices for each $K_0^\text{ch}$. 
Fig.~\ref{fig:svd_analysis}(b) shows the decay of the singular values for each $K_0^\text{ch}$. 
As previously noted, the singular values of $K_0^\text{off-ch}$ decay much faster than the $K_0^\text{diag-ch}$.
Fig.~\ref{fig:svd_analysis}(c) shows the minimum ratio of singular values $z'/N_k$ required to reconstruct each $K_0^\text{ch}$ with a high accuracy of $R^2=0.98$ ($R^2=1-\Sigma_i(y_i-\hat{y}_i)^2/\Sigma_i(y_i-\bar{y}_i)^2$, where $y_i$ and $\hat{y}_i$ are real and predicted values).
The $K_0^\text{diag-ch}$ requires a high SVD truncation ratio of 95\% for reconstruction, while the mean truncation ratio for the $K_0^\text{off-ch}$ is only $\mathrm{12\%}$. Fig.~\ref{fig:svd_analysis}(d) and (e) respectively show the channel-wise truncation rate of the direct $K_0^\text{d}$ and exchange $K_0^\text{x}$ interaction only. Notably, the exchange interaction
is highly compressible, requiring a truncation ratio of only 2\%. The direct term alone is less compressible with a mean truncation ratio of 14\% for off-diagonal channels.

Eqs.~\eqref{eq:svd_diag_offdiag} and \eqref{eq:svd-channel-wise} can both be utilized to compress the BSE kernel and accelerate TD-aGW simulations.
As the number of bands increases, the SVD compression scheme described in Eq.~\eqref{eq:svd_diag_offdiag} poses challenges for both calculations, which scale as $\mathcal{O}((N_{\text{b}}^2N_{\vb{k}})^3)$, and memory. Alternatively, the time complexity of channel-wise SVD (Eq.~\ref{eq:svd-channel-wise}) is $\mathcal{O}(N_{\text{b}}^4(N_{\vb{k}})^3)$, and a high overall SVD compression rate can still be achieved as non-compressible diagonal channels only account for $1/N_{\text{b}}^2$ of $K_0$. Moreover, channel-wise SVD is inherently well-suited for parallel implementation and scalability.

Lastly, we investigate the SVD compression for different sizes of the k-grid. Fig.~\ref{fig:svd_analysis}(f) shows the average minimum number of $z$ required to reconstruct $K_0^\text{off-ch}$ for varying k-grid sizes. Notably, for a given accuracy requirement, the number of $z$ tends to converge as the size of the k-grid increases, indicating that additional k-points do not proportionally contribute new information to the $K_0$. 
This key observation ensures computational feasibility even when an extremely dense k-grid is required for BSE and TD-aGW, where k-grid convergence is a major bottleneck. 
Moreover, since $z$ remains independent of $N_{\vb{k}}$, the compression ratio $z/N_{\text{b}}^2 N_{\vb{k}}$ will decrease with increasing $N_{\vb{k}}$, significantly accelerating calculations with large $\vb{k}$-grids.

\subsection{SVD acceleration of TD-aGW}\label{sec:svd-tdagw}

\begin{figure}
    \centering
    \includegraphics[width=\linewidth]{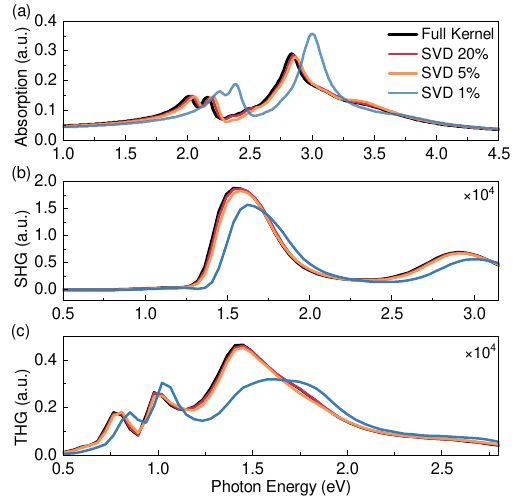}
    \caption{
        (a) The optical absorption spectrum of monolayer $\mathrm{MoS_2}$ in arbitrary units (a.u.) is calculated in TD-aGW linear response after excitation by a narrow Gaussian electric pulse. The TD-aGW simulations are conducted with the full $K_0$ (dark) and the $\Tilde{K}_0$ that includes 20\% (red), 5\% (orange), and 1\% (blue) of singular values. The (b) second- and (c) third-order response intensity of monolayer $\mathrm{MoS_2}$ to incident electric field frequencies from 0.4 to 3.2 eV using full $K_0$ and the $\Tilde{K}_0$. The external field is a sine wave with a field amplitude of $2.36 \times 10^7$ V/cm$^2$.  
    }
    \label{fig:abs_linear}
\end{figure}

Next, we apply the SVD-compressed kernel to accelerate TD-aGW calculations. In Eq.~\eqref{eq:tdagw}, the time-dependent Hamiltonian $H(t)=H^\text{QP}-e\mathbf{E}(t) \cdot \mathbf{r} + \delta \Sigma(t)$ 
is updated at each time step. 
The major computational bottleneck lies in evaluating the change in the COHSEX self-energy $\delta \Sigma(t)$ (Eq.~\eqref{eq:sigma-from-k}), which has a complexity of  
$\mathcal{O}(N_{\text{b}}^4N_{\vb{k}}^2)$ per time step. Further analysis of the time complexity is given in Appendix A.
We can reduce this complexity by taking advantage of the low-rank structure of $K^\text{off}_0$. We plug the SVD truncated kernel (Eq.~\ref{eq:svd_diag_offdiag}) into the TD-aGW Hamiltonian:

\begin{equation}
\begin{aligned}
\Tilde{H}(t) &\approx H^\text{QP} - e\mathbf{E}(t) \cdot \mathbf{r} + K^{\text{diag}}_0 \rho(t) + \Tilde{U}_0\Tilde{M}_0\Tilde{V}_0^T \rho(t)  \\
&= H^\text{diag}(t) + \Tilde{S}_0(\Tilde{V}_0^T \rho(t))
\label{eq:svd_acc}
\end{aligned}
\end{equation}
Here, the diagonal part of Hamiltonian $H^\text{diag}(t) =H^\text{QP}+ K^{\text{diag}}_0 \rho(t) - e\mathbf{E}(t) \cdot \mathbf{r}$ is time-dependent but can be evaluated in a negligible $\mathcal{O}(N_{\text{b}}^2N_{\vb{k}})$. $\Tilde{S}_0=\Tilde{U}_0\Tilde{M}_0$ is time-independent, so the acceleration is achieved by first projecting the density matrix $\rho(t)$ onto the low-rank subspace spanned by the truncated singular vectors $\Tilde{V}_0^{T}$, then mapping it back to the full Hilbert space via $\Tilde{S}_0$. The time complexity of this term is reduced to $\mathcal{O}(N_{\text{b}}^2N_{\vb{k}} \times 2z)$, accelerating TD-aGW by a factor of $N_{\text{b}}^2N_{\vb{k}}/2z$ compared to the full kernel case.
We note that with a given accuracy, $z$ is almost independent of $N_{\vb{k}}$ in the large $N_{\vb{k}}$ limit, and therefore the SVD-compression approach is particularly powerful with a dense k-grid, which is needed for convergence of the BSE kernel \cite{qiu2013optical}.

As an initial benchmark, we validate the SVD-accelerated approach by comparing the linear optical absorption spectrum obtained from TD-aGW using full $K_0$ to $\Tilde{K}_0$ with different SVD truncation ratios. We conduct TD-aGW simulations for 150 fs with a time step of 0.125 fs, and a narrow Gaussian pulse of width 0.125 fs centered at 0.125 fs  is applied to excite the system. The absorption spectrum $A(\omega)$ is obtained by a Fourier transformation of the time-dependent polarization $P(t)$\cite{chan2021giant}.
Fig.~\ref{fig:abs_linear} (a) shows the linear absorption spectrum of monolayer $\mathrm{MoS_2}$ calculated with the full $K_0$ and SVD subspace truncated to include 20\%, 5\% and 1\% of singular values. Notably, the absorption spectrum with a 5\% SVD subspace (resulting in a $10\times$ speedup) can accurately reproduce the result from the full calculation with less than 0.04 eV error in the position of the first exciton peak. When the SVD truncation threshold is decreased to 1\% of singular values, there is a significant blue shift of the absorption spectrum, which arises from the information loss in the direct Coulomb interaction in the kernel.

The SVD-accelerated TD-aGW method is robust to different setups and parameters of the external driving field due to the preservation of the main physical interaction. To demonstrate this, we perform a parameter sweeping of the incident field frequency $\omega$. Fig.~\ref{fig:abs_linear} (b-c) shows the second harmonic generation (SHG) and the third harmonic generation (THG) intensity of monolayer $\mathrm{MoS_2}$ corresponding to plane-wave incident light with frequencies ranging from 0.4 to 3.2 eV with an interval of 0.04 eV. As we expect, the SVD-accelerated TD-aGW with 5\% truncation is still enough to capture both SHG and THG spectrum patterns, while the 1\% truncation kernel results in an obvious deviation from the ground truth, in line with the linear response results.

\begin{figure}
    \centering
    \includegraphics[width=\linewidth]{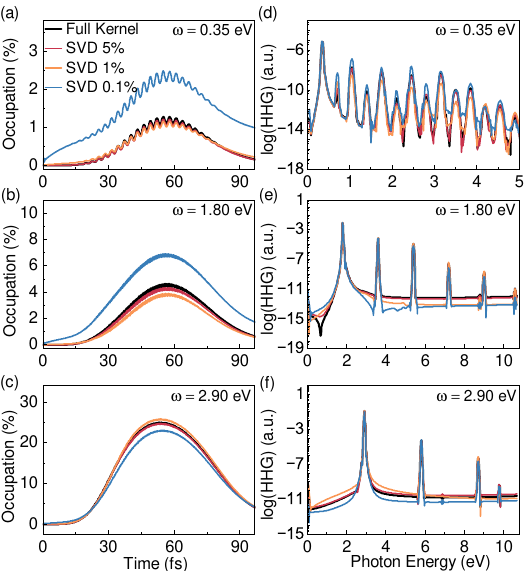}
    \caption{TD-aGW simulations in the strong field regime with the full $K_0$ (black) vs. SVD subspace with different singular value truncation ratios of 5\% (red), 1\% (orange), and 0.1\% (blue). (a-c) Time-dependent electronic occupation of the first conduction band $\mathrm{c_1}$ in MoS$_2$ with driving field frequencies of (a) 0.35, (b) 1.80, and (c) 2.90 eV. The occupation percentage is computed as $\frac{1}{N_{\vb{k}}} \sum_{\vb{k}} \rho_{c_1,c_1,\vb{k}}(t)$, where $N_{\vb{k}}$ is the number of k points. (d-f) The high-harmonic generation (HHG) spectrum for the same driving field energies as (a-c), respectively.}
    \label{fig:svd_time_freq}
\end{figure}

Next, to evaluate the performance of SVD-accelerated TD-aGW 
in the nonperturbative strong field regime, we conduct simulations with the external field used in previous nonlinear optical experiments \cite{liu2017high} and calculations \cite{chang2024many}:
$\mathbf{E(t)} = \mathbf{E_0} \cdot  \sin^2\left(\pi  tT_{\text{pulse}}\right) \cdot \sin(\omega t)$
, where the electric field amplitude $\mathbf{E_0}$ is \SI{2.36e7}{V/cm^2},  $T_{\text{pulse}}$ is 100 fs, and $\omega$ is the incident pulse frequency.  The simulation length is 100 fs with a time step of $\dd{t}=\SI{0.1}{fs}$. Fig.~\ref{fig:svd_time_freq}(a-c) shows the time-dependent occupation $ \sum_{\vb{k}} \rho_{nn\vb{k}}(t) / N_{\vb{k}}$ of the first conduction band $c_1$ in monolayer $\mathrm{MoS_2}$ with excitation field energies in the gap (0.35 eV), resonant with the lowest bright exciton (1.80 eV), and above the continuum (2.90 eV). For the in-gap excitation (Fig.~\ref{fig:svd_time_freq}(a)), the maximum excited-state population fills less than 1.5\% of the conduction band, while the occupation increases to 25\% for excitation above the continuum (Fig.~\ref{fig:svd_time_freq}(c)). For below-gap excitations, a SVD truncation ratio of 1\% (50$\times$ speedup) is sufficient to accurately capture both the occupation and harmonic yield.
When the SVD truncation ratio is lowered to 0.1\%, a substantial deviation from the ground truth is observed, indicating that the $K_0^\text{off}$ is not negligible even if it is much smaller than the $K_0^\text{diag}$. However, as the excitation energy increases(Fig.~\ref{fig:svd_time_freq}(b-c)), the numerical deviation between the 0.1\% SVD-accelerated TD-aGW from the ground truth becomes smaller. 
A possible explanation is that $K_{0}^\text{off}$ is more important for in-gap excitations, where the small energy of the incident photon can drive transitions between different valence bands or conduction bands, where as the larger dirving fields are resonant with specific valence to conduction band transitions.

Importantly, compared to previous data-driven methods and operator learning approaches for solving the KBE \cite{gu2024probabilistic, reeves2023dynamic, reeves2024performance}, our SVD-accelerated TD-aGW avoids the time-accumulated errors that quickly cause the prediction to diverge from the ground truth. This arises from two key factors: i) the SVD-accelerated TD-aGW leverages the linear low-rank structure of the kernel matrix to retain most of the physical couplings for non-equilibrium dynamics; and ii) our SVD compression is independent of the external driving field, whereas supervised learning methods may struggle to capture dynamic behaviors driven by different external fields that deviate significantly from the training set.

To demonstrate the effectiveness of SVD-accelerated TD-aGW in the frequency domain, we calculate the high-harmonic generation (HHG) spectrum by $\text{HHG}(\omega) = \left| \omega \int \mathbf{J}(t) \ee^{-\ii \omega t} \dd{t} \right|^2$\cite{chang2024many}
, where the time-dependent current is calculated by $\mathbf{J}(t)=\Tr(\rho(t) \mathbf{v})$ and $\mathbf{v}$ is the velocity operator. Fig.~\ref{fig:svd_time_freq} (d) shows the logarithmic magnitude of the HHG spectrum with an incident pulse frequency of $\omega=$ 0.35 eV. The HHG calculation using 5\% truncated kernel agrees well with the ground truth. The 0.1\% truncation can only reproduce the intensity of the 1st-order response, but the relative strength of the high harmonics are still faithfully reproduced. As shown in Fig.~\ref{fig:svd_time_freq}(e-f), when the incident photon frequencies are increased to 1.8 and 2.9 eV, even the low truncation ratio of 0.1\%  can still agree well with the ground truth for the high-harmonic generation due to the decreased contribution from the $K_0^\text{off}$ in the high-frequency region of the incident pulse. We note, however, that while the above gap pump is an informative numerical test, it does not reflect physically realistic conditions, as the pump intensity required to produce high harmonics would burn the sample. For the parameter sweeping of the angle-dependent HHG spectrum see Appendix C.

\subsection{SVD compression for GW-BSE within TDA}\label{sec:absorption}
    
\begin{figure}
    \centering
    \includegraphics[width=\linewidth]{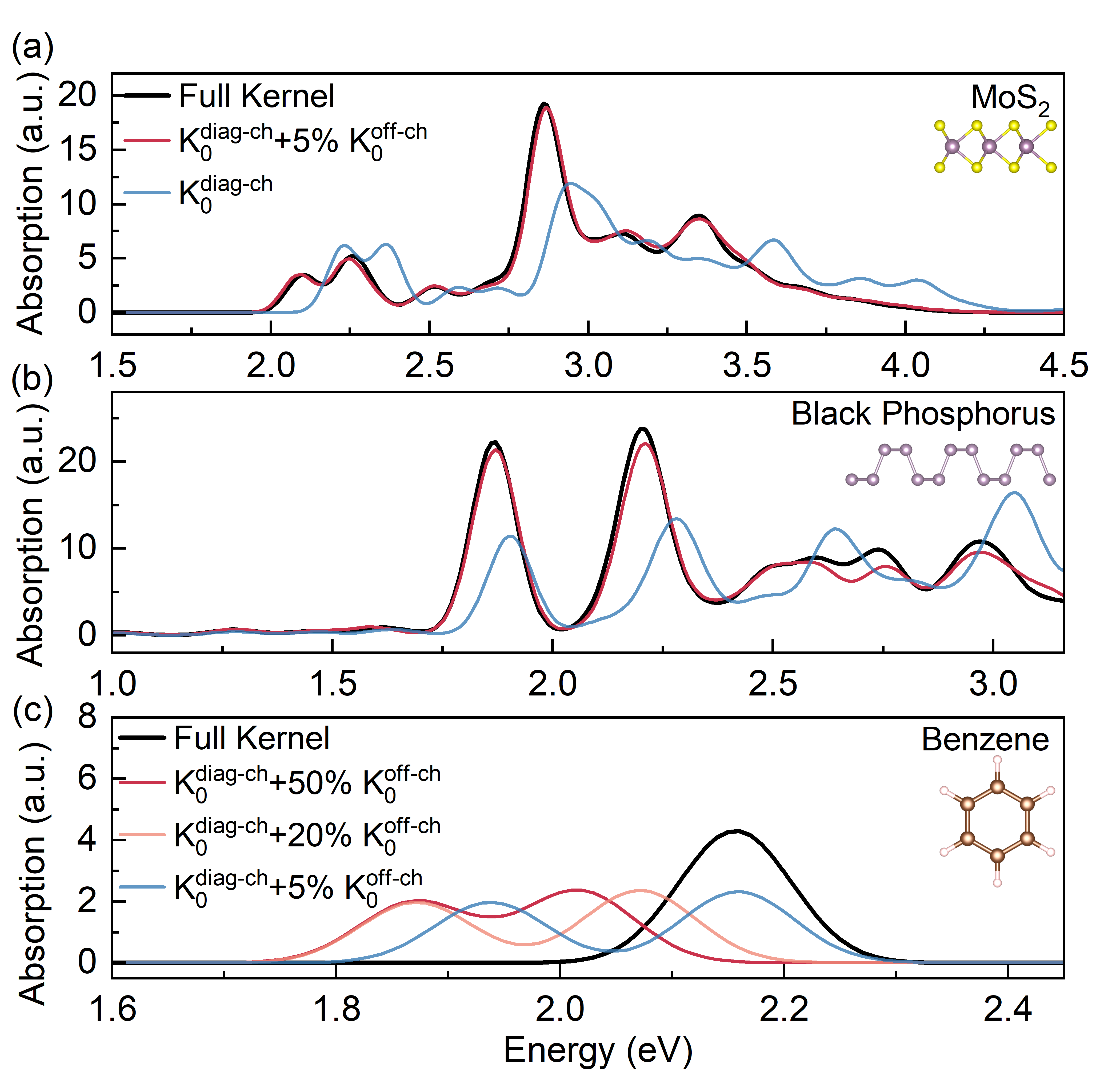}
    \caption{(a-c) The absorption spectrum calculated through the equilibrium GW-BSE formalism of (a) monolayer \ce{MoS2}, (b) monolayer black phosphorus, and (c) a Benzene molecule. In (a-b) SVD compression is conducted over the momentum space $\ket{\vb{k}}$ of $K_0^\text{ch}$ for crystal systems. The BSE is calculated with the full $K_0$(black), $K_0^\text{diag-ch}$+5\%$K_0^\text{off-ch}$(red), and $K_0^\text{diag-ch}$ only (blue). (c) SVD compression is conducted over the band-to-band transition channels $\ket{n_1,n_2}$ of the molecule system due to lack of periodicity. The BSE is calculated with the full $K_0$(black), $K_0^\text{diag-ch}$+50\% $K_0^\text{off-ch}$(red), +20\% $K_0^\text{off-ch}$(orange), +5\% $K_0^\text{off-ch}$(blue)
    }
    \label{fig:tdapprox_BSE}
\end{figure}

Up to now, we have used monolayer MoS$_2$ as a testbed for SVD compression. Here, we explore material dependence 
and demonstrate that the optical absorption can be reliably calculated with $\Tilde{K}_0$ across extended crystalline materials.
In \prettyref{fig:tdapprox_BSE}(a-b), we apply channel-wise SVD to the GW-BSE kernel in the Tamm-Dancoff approximation (TDA) for monolayer \ce{MoS2} and black phosphorus, which are both 2D semiconductors with exciton binding energies on the order of several hundred meVs and exciton wavefunctions confined to low-energy valleys in $\vb{k}$-space\cite{Qiu2012MoS2, cao2016gate, qiu2017environmental, li2017direct}. Calculation details are given in Appendix A.
As Fig.~\ref{fig:tdapprox_BSE}(a-b) illustrates, electron-hole kernel matrices of \ce{MoS2} and black phosphorus exhibit similar low-rank structure, with an average 5\% SVD compression of $K_0^\text{off}$ being sufficient to accurately reproduce the linear absorption spectra. 

We have studied electron-hole interactions in extended systems, where we perform a channel-wise SVD compression over $\vb{k}$-points. 
Lastly, we consider a benzene molecule in the gas phase, which lacks periodicity but converges slowly with respect to the number of bands~\cite{Qiu2012Sapprox}.
Here, without the crystal momentum $\vb{k}$ degree-of-freedom, the SVD can be only applied to the transition channels $\ket{n_1,n_2}$. However, even with a modest compression rate of 50\%, compression over bands introduces unphysical peak splitting, highlighting that while SVD compression works well over $\vb{k}$-points, it is fundamentally unsuitable for molecular systems that lack periodicity.

Finally, we  note that SVD compression can be trivially generalized to a finite momentum BSE kernel (Appendix D), suggesting a route for further accelerating dynamcis calculations simulating scattering between finite momentum states.

\section{Conclusion}\label{sec:conclusion}

In summary, we reveal a low-rank structure of the electron-hole kernel, which arises due to the localization of excitonic effects in momentum space in crystalline systems. As a consequence of the low-rank structure, we show that the full kernel matrix can be compressed to at least 5\% of its original size in semiconductors with exciton binding energies on the order of a few hundreds of meV.
Notably, we show that the size of subspace with non-zero singular values, capturing key information about the electron-hole kernel, tends to converge as the $\vb{k}$-grid size increases, ensuring computational feasability even for dense $\vb{k}$-grids.
Leveraging this general low-rank property, we apply it to accelerate non-equilibrium TD-aGW calculations, achieving a speedup of an order of magnitude in calculating the time-dependent density matrix, angle-resolved HHG in the strong-field regime, and optical absorption spectrum in the weak-field limit for monolayer \ce{MoS2}. Here, we demonstrate SVD acceleration in the context of zero-momentum excitations in TD-aGW, but our formalism can be trivially generalized to finite momentum BSE calculations, where we show that 5\% of SVD kernel data can generate electron loss spectroscopy comparable to the ground truth (see Appendix D).
Our method avoids time-accumulated prediction errors seen in  data-driven and operator learning methods for dynamic calculation of nonequilibrium many-body effects. It is also requires no training data and is agnostic to the choice of crystal systema and external driving field, regardless of whether the driving field lies above or below the bandgap or in the weak or strong-field regime. 
Our approach provides a robust, efficient, and broadly applicable framework for accelerating many-body calculations, offering new possibilities for studying nonequilibrium excitonic and electronic dynamics with high accuracy and reduced computational cost.

\section{Appendix A: Computational Details}

Firstly, we note that the $\vb{G}=0$ contribution in Eq.~\eqref{eq:bse-exchange-def}, which corresponds to the long-range, macroscopic Coulomb interaction, may be zeroed out in actual simulations so that the $\vb{E}$ field in Eq.~\eqref{eq:tdagw} is the total macroscopic electric field and not just the external electric field.
This means the linear and nonlinear response from Eq.~\eqref{eq:tdagw} are calculated with respect to the total macroscopic electric field, which is consistent with the way material-specific constitutive relations are defined in electrodynamics in media.

In this work, all DFT calculations are performed using Quantum ESPRESSO\cite{Giannozzi_2009} with Perdew-Burke-Ernzerhof (PBE) functional\cite{PBE1}, and all GW-BSE calculations are performed using BerkeleyGW\cite{BGW1,BGW2,BGW3}. For zero-momentum GW-BSE calculations of $\mathrm{MoS_2}$, the cutoff of DFT kinetic energy, RPA dielectric matrix, and kernel matrix are 95, 20, and 5 Ry respectively. 6,000 bands are included in constructing the dielectric matrix. The $K_0$ of $\mathrm{MoS_2}$ includes 2 conduction and 2 valence bands, which are sampled on a uniform k-grid of $36\times36\times1$ over the first Brillouin zone. For black phosphorus, the cutoff of DFT kinetic energy, RPA dielectric matrix, and kernel matrix are 95, 20, and 5 Ry respectively. 1,000 bands are included in constructing the dielectric matrix. The $K_0$ of black phosphorus includes 4 valence and 4 conduction bands, which are sampled on a k-grid of $20\times20\times1$. For the benzene molecule, the DFT plane-wave cutoff energy is 90 Ry. 600 bands are included in evaluating the dielectric matrix and kernel matrix with a cutoff energy of 40 and 10 Ry respectively.

\subsection{A.1: Time-complexity of TD-aGW}

Here, we analyze the time complexity of Eq.~\eqref{eq:tdagw}, where the time-dependent Hamiltonian $H(t)=H^\text{QP}-e\mathbf{E}(t) \cdot \mathbf{r} + \delta \Sigma(t)$ 
is updated at each time step. 
The quasiparticle Hamiltonian $H^\text{QP} \in \mathbb{R}^{N_{\text{b}}^2 N_{\vb{k}}}$ is time-independent and calculated through DFT mean-field calculations with many-body $G_0W_0$ corrections prior to TD-aGW.
The dipole matrix $\mathbf{r} \in \mathbb{C}^{N_{\text{b}}^2 N_{\vb{k}} \times 3}$ is obtained after transforming wavefunctions to a locally smooth gauge equivalent to the parallel transport gauge \cite{localgauge1,localgauge2,localgauge3,localgauge4}. This matrix does not change during the time evolution so the external field term can be easily updated by taking $e\mathbf{E}(t) \cdot \mathbf{r}$, which scales as a negligible $\mathcal{O}(N_{\text{b}}^2N_{\vb{k}})$. The major computational bottleneck lies in evaluating the change in the COHSEX self-energy $\delta \Sigma(t)$ (Eq.~\eqref{eq:sigma-from-k}), and its time complexity scales as 
$\mathcal{O}(N_{\text{b}}^4N_{\vb{k}}^2)$ per time step.

\section{Appendix B: Analysis of Electron-hole kernel}

\begin{figure}
    \centering
    \includegraphics[width=\linewidth]{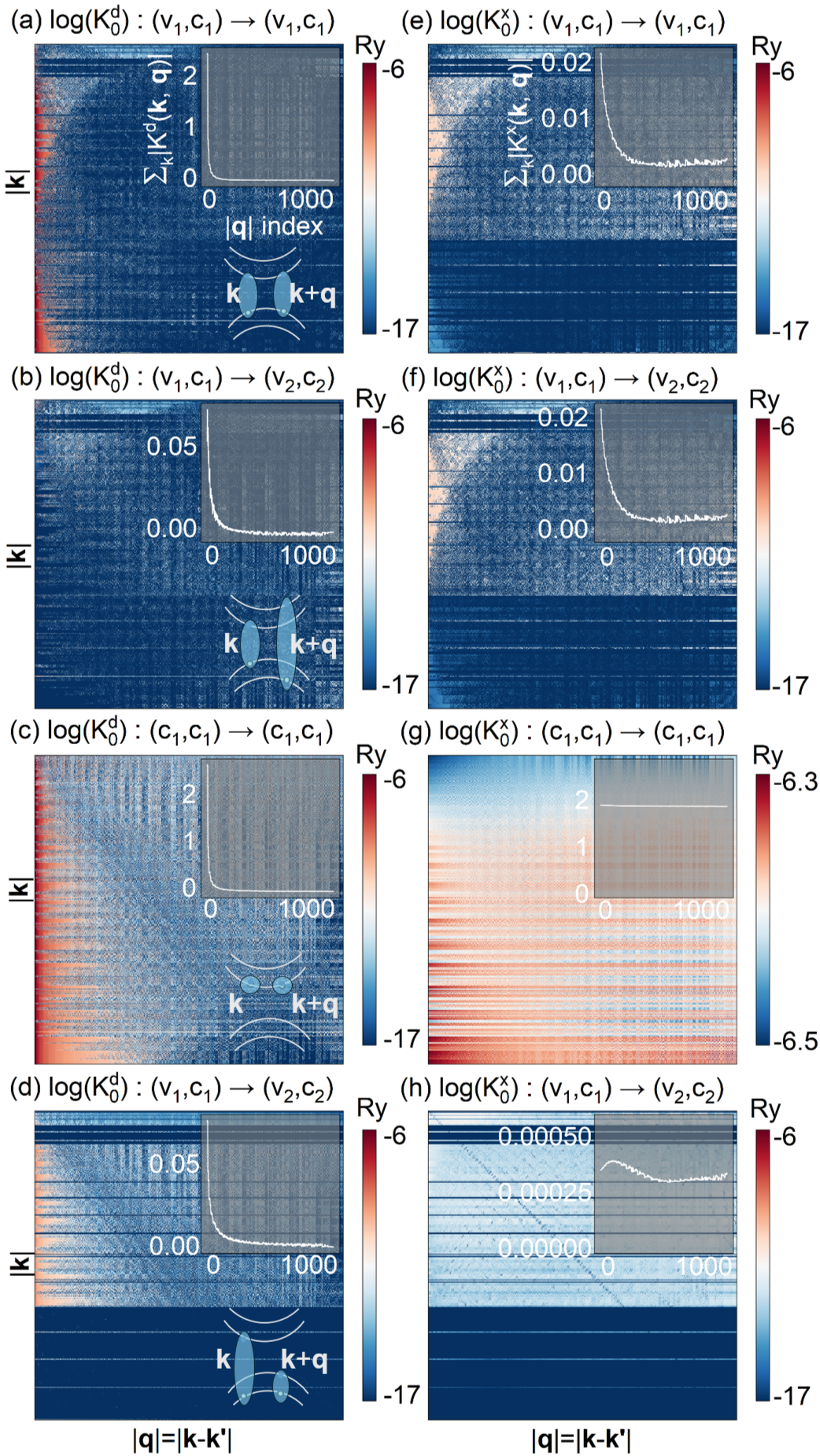}
    \caption{Structure of the electron-hole kernel $K_0$ of monolayer \ce{MoS2}. To visualize the $|\vb{k} - \vb{k}'|$ dependence of direct $K_0^d$ and exchange $K_0^x$, the kernel matrix elements $K_0(\vb{k}, \vb{k}')$ are reordered to $K_0(|\vb{k}|, |\vb{q}|)$, where $\vb{q} = \vb{k} - \vb{k}'$. The color map shows the log of the magnitude of $K_0$.  (a-b) The $K_0^d$ between electron-hole pairs across valence-to-conduction transition channels of $\langle v_1 c_1, \vb{k} | K_0^d | v_1 c_1, \vb{k} + \vb{q} \rangle$ and $\langle v_1 c_1\vb{k} | K_0^d | v_2 c_2\vb{k} + \vb{q} \rangle$, which are typically used in equilibrium BSE. (c-d) the $K_0^d$ involving electron-hole channel within the conduction or valence bands $\langle c_1 c_1\vb{k} | K_0^d | c_1 c_1\vb{k} + \vb{q} \rangle$ and $\langle v_2 c_1\vb{k} | K_0^d | v_2 v_1\vb{k} + \vb{q} \rangle$, which are crucial for nonequilibrium dynamics in TD-aGW. (e-h) represent the $K_0^x$ for channels corresponding to (a-d). The inset curves illustrate the decay of kernel interaction with respect to $|\vb{q}|$, and the inset schematics illustrate the electro-hole interaction channel in each plot.}
    \label{fig:kernel_overview}
\end{figure}

Fig.~\ref{fig:kernel_overview} is a logscale plot of the magnitude of the direct ($K_0^\text{d}$) and exchange ($K_0^\text{x}$) kernel matrix elements plotted over $\vb{k}$ and $\vb{q = k-k'}$ for selected channels.
Fig.~\ref{fig:kernel_overview} (a) and (b) depict the direct Kernel for a diagonal channel $(v_1, c_1) \to (v_1, c_1)$ and off-diagonal channel $(v_1, c_1) \to (v_2, c_2)$, respectively. These electron-hole pairs are purely composed of valence-to-conduction transitions, which make up the equilibrium BSE Hamiltonian in the Tamm-Dancoff approximation (TDA)\cite{Rohlfing2000}. When $\vb{k}=\vb{k}'$ or $\vb{q}=0$, the screened Coulomb interaction diverges. To make the computation numerically tractable, we replace $W(\vb{q}=0)$ with an average over a small region of the Brillouin zone~\cite{deslippe2012berkeleygw}. However, even when the divergent term is removed, the magnitude of the direct term in $K_0^\text{diag-ch}$ increases exponentially at $\vb{q}=0$ across all $\vb{k}$-vectors because of the large wave function overlap (\prettyref{fig:tdagw-diagram}(b), Eq. \eqref{eq:bse-direct-def}), suggesting that the diagonal channels of the direct kernel are likely to be full-rank matrices. 
In contrast, the $K_0^\text{off-ch}$ (Fig.~\ref{fig:kernel_overview}(b)) as well as the non-diagonal (i.e. $\vb{k} \neq \vb{k}'$) part of $K_0^{\text{diag-ch}}$ are sparse across both $\vb{k}$ and $\vb{q}$ and are likely to be low-rank. 
This is consistent with and also explains the previous observation that exciton wave functions tend to be confined in small patches in 1BZ \cite{alvertis2023importance, Hou2023Bi2Se3,li2017direct,Wu2024WTe2, Qiu2012MoS2}.
Fig.~\ref{fig:kernel_overview} (c) and (d) show the structure of the direct kernel for the diagonal $(c_1, c_1) \to (c_1, c_1)$ and off-diagonal $(v_2, c_1) \to (v_2, v_1)$ channel, which do not contribute to the equilibrium BSE but are important for understanding nonequilibrium response in TD-aGW. 
They behave similarly to the TDA channels shown in \prettyref{fig:kernel_overview}(a) and (b), with the diagonal channel being close to full rank and the off-diagonal channel being sparser.

Fig.~\ref{fig:kernel_overview} (e-h) show the same channels as Fig.~\ref{fig:kernel_overview} (a-d) for the exchange interaction instead of the direct interaction. 
For the excitation channels contributing to the TDA kernel (e-f), the inset curve shows that the exchange interaction is smaller than the direct interaction.
For the beyond-TDA channels, (g-h), the diagonal channel $(c_1, c_1) \to (c_1, c_1)$ has a very large exchange interaction (due to the large overlap of $c_1$ with $c_1$; \prettyref{fig:tdagw-diagram}(b), Eq. \eqref{eq:bse-exchange-def}) that is almost constant with respect to $\vb{q}$. The near-constant exchange arises from the fact that the Coulomb interaction in the exchange only explicitly depends on the exciton center-of-mass momentum $\vb{Q}$ instead of the momentum transfer $\vb{q}$ (\prettyref{fig:tdagw-diagram}(b), Eq. \eqref{eq:bse-exchange-def}). A matrix with multiple constant rows or columns is also expected to be low rank.

\section{Appendix C: Angle-resolved HHG}

\begin{figure}
    \centering
    \includegraphics[width=\linewidth]{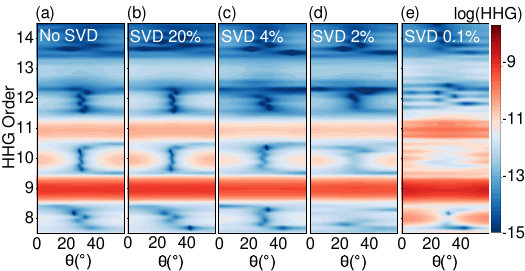}
    \caption{(a-e) represent the angle-resolved HHG computed within TD-aGW using the full kernel, and SVD subspaces of 20\%, 4\%, 2\%, and 0.1\%, respectively. The angle incident $\theta$ ranges from 1 to 60$^{\circ}$ with an interval of 1$^{\circ}$, which depicts the polarization of the driving field oriented at an angle $\theta$ with respect to the mirror plane (110). 
}
    \label{fig:high_order_response}
\end{figure}

To demonstrate the SVD effect on parameter sweeping, we also calculate the angle-resolved HHG spectrum of different incident polarization angles $\theta$ from 0$^{\circ}$ to 60$^{\circ}$ with an interval of 1$^{\circ}$. 
The Fig.~\ref{fig:high_order_response} (a-d) shows that the SVD-accelerated TD-aGW with 2\% truncation can accurately capture the symmetry of the HHG spectrum, while the 0.1\% truncation kernel results in a significant deviation from the ground truth.

\section{Appendix D: EELS}
We show here that SVD compression can be extended to the electron-hole kernel at finite momenta. We focus on the electron energy loss spectra (EELS) calculated in the GW-BSE approach within the TDA ad the main observable. In the zero-momentum limit, when the long-range exchange term $K_{\text{x}}(\vb{q}, \vb{G}=0)$ in BSE is zeroed out, the BSE only describes transverse excitons \cite{qiu2021signatures} and corresponds to the irreducible response function to the external macroscopic electromagnetic field \cite{ambegaokar1960electromagnetic,gatti2013exciton}.
The solutions of the BSE can therefore be utilized to calculate the imaginary part of the macroscopic transverse dielectric function via  Fermi's golden rule in Rydberg units \cite{ambegaokar1960electromagnetic,gatti2013exciton,BGW1}
\begin{equation}
\begin{aligned}
    &\Im \epsilon_\text{M} (\textbf{q} = 0, \omega) = \\ &8 \pi^2 \sum_{S} \abs*{\vu{e} \cdot \sum_{vc\textbf{k}}A^S_{vc\textbf{k}}\bra{v\textbf{k}} \vb{r} \ket{c\textbf{k}}}^2 \delta(\omega - \Omega^S),
\end{aligned}
\label{eq:eps}
\end{equation}
which gives the absorption spectrum.
On the other hand, finite momentum BSE calculations containing the long-range exchange term correspond to the reducible response function to the external macroscopic electromagnetic field \cite{ambegaokar1960electromagnetic,gatti2013exciton}.
Therefore a finite momentum BSE calculation at $\vb{q}$ can be utilized to calculate the imaginary part of the inverse of the longitudinal macroscopic dielectric function \cite{gatti2013exciton}
\begin{equation}
\begin{aligned}
    &-\Im \frac{1}{\epsilon_\text{M} (\textbf{q}, \omega)} = \\
    &\quad \frac{8 \pi^2}{q^2} \sum_{S}| \sum_{vc\textbf{k}}A^{S \vb{Q}}_{vc\textbf{k}}\bra{v \vb{k} - \vb{q}} \ee^{\ii \vb{q} \cdot \vb{r}}\ket{c\textbf{k}}|^2 \delta(\omega - \Omega^S),
\end{aligned}
\label{eq:eps-inv}
\end{equation}
where the momentum $\vb{Q}$ of excitons contributing to $\Im \epsilon_{\text{M}}^{-1}(\vb{q}, \omega)$ at $\vb{q}$ is the same as $\vb{q}$.
The inverse of the longitudinal macroscopic dielectric function is proportional to the electron energy loss spectrum \cite{egerton2008electron,kuzmany2009solid}.

Now we can study effects of SVD compression for the center-of-mass momentum-dependent kernel, $K_{\vb{Q}}$ by performing finite-momentum BSE calculation on monolayer \ce{MoS2} at each exciton momentum $\vb{Q}$ with full and SVD reconstructed $\Tilde{K}_{\vb{Q}}$ and compare the EEL spectra obtained from the two approaches at each $\vb{Q}$. 
For the finite-momentum BSE calculation, DFT kinetic energy cutoff is 95 Ry. The cutoff for the RPA dielectric matrix and kernel are 20 and 5 Ry. The finite-momentum kernel matrix includes 2 conduction bands and 2 valence bands over a $24 \times 24 \times 1$ $\vb{k}$-grid.
We then calculate the electron energy loss spectrum based on Eq. \eqref{eq:eps-inv}.

\begin{figure}
    \centering
    \includegraphics[width=1.0\linewidth]{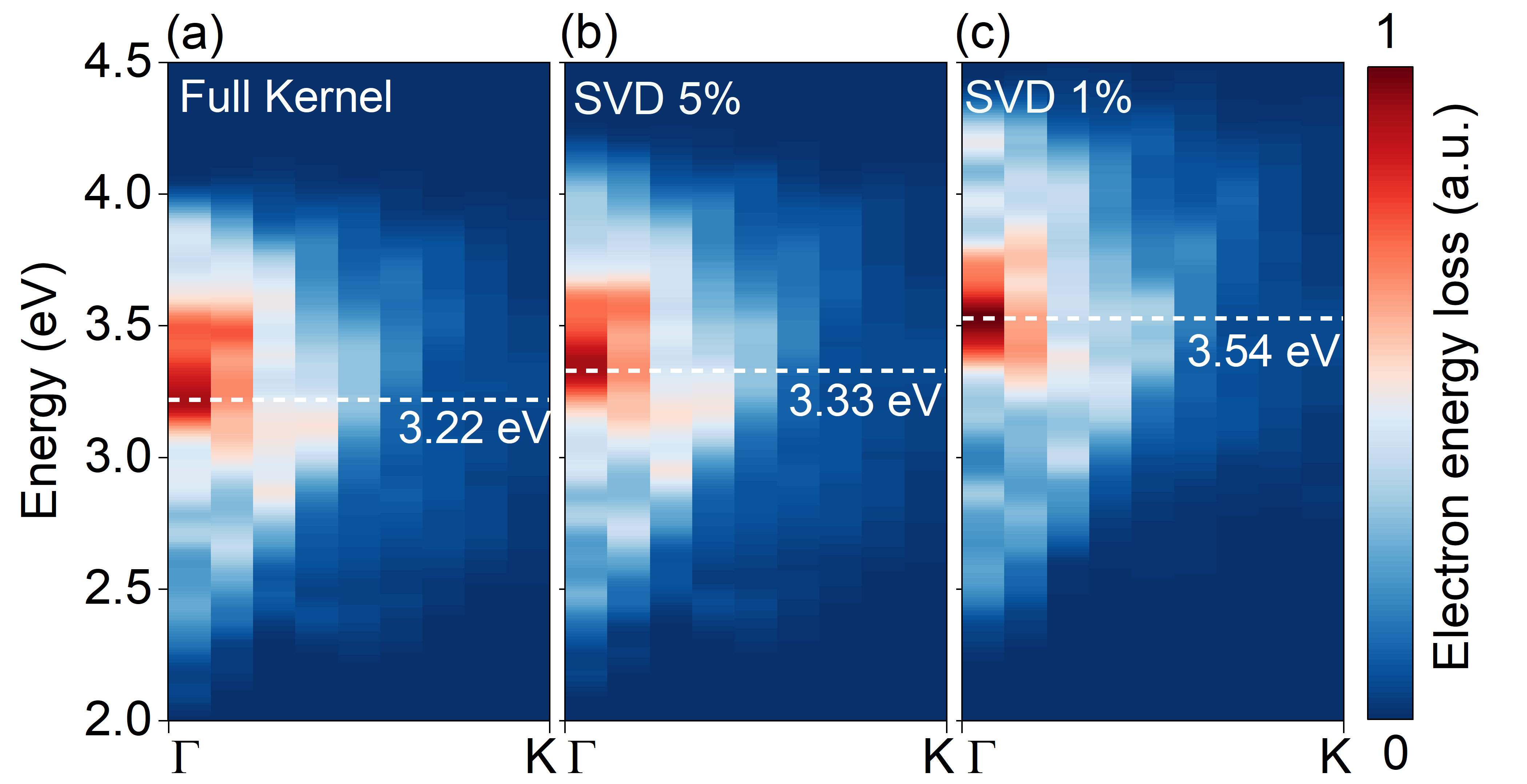}
    \caption{Electron-energy loss spectroscopy simulation based on finite momentum BSE along the $\Gamma$-K $\vb{k}$-path in monolayer 1T' \ce{MoS2}. The BSE kernel used in calculating the right panel undergoes an SVD compression that only keeps 5\% of the original singular values.
    Gaussian broadening with a width of \SI{0.05}{eV} is applied to each mode.
    The center of the brightest peak in the EELS is marked by a dashed white line as a guide to the eye.}
    \label{fig:eels-K-G}
\end{figure}

The result in \prettyref{fig:eels-K-G} shows that  EELS obtained from the full $K_Q$ and the kernel with 5\% SVD truncation are similar each other, with clear correspondence between the peaks in the two heatmaps.
Quantitatively, the $R^2$ between the $\Im \epsilon_{\text{M}}^{-1}(\vb{q}, \omega)$ (Eq.~\ref{eq:eps-inv}) before and after the \SI{5}{\percent} SVD compression is around $\sim 0.91$.
The $R^2$ value (defined in \prettyref{sec:bse-kernel-svd}) drops to $\sim 0.62$ when only \SI{1}{\percent} of the singular values are kept,
and the structure of the electron energy loss spectrum also shows more deviation from the non-compressed spectrum.
We also note that the peaks in the electron energy loss spectra show a blueshift as we compress the BSE kernels, which is intuitive because the loss of information in $K_{\vb{Q}}$ pushes the BSE calculation towards the non-interacting limit.

\section{Acknowledgement}
This work was primarily supported by the U.S. Department of Energy, Office of Science, Basic Energy Sciences under Early Career Award No. DE-SC0021965. Development of the BerkeleyGW code was supported by Center for Computational Study of Excited-State Phenomena in Energy Materials (C2SEPEM) at the Lawrence Berkeley National Laboratory, funded by the U.S. Department of Energy, Office of Science, Basic Energy Sciences, Materials Sciences and Engineering Division, under Contract No. DE-C02-05CH11231. Calculations on benzene and black phosphorus were supported by the National Science Foundation Division of Chemistry under award number CHE-2412412. The calculations used resources of the National Energy Research Scientific Computing (NERSC), a DOE Office of Science User Facility operated under contract no. DE-AC02-05CH11231; the Advanced
Cyberinfrastructure Coordination Ecosystem: Services \& Support (ACCESS), which is
supported by National Science Foundation grant number ACI-1548562; and the Texas Advanced Computing Center (TACC) at The University of Texas at Austin.

\bibliography{main}

\begin{thebibliography}{93}%
\makeatletter
\providecommand \@ifxundefined [1]{%
 \@ifx{#1\undefined}
}%
\providecommand \@ifnum [1]{%
 \ifnum #1\expandafter \@firstoftwo
 \else \expandafter \@secondoftwo
 \fi
}%
\providecommand \@ifx [1]{%
 \ifx #1\expandafter \@firstoftwo
 \else \expandafter \@secondoftwo
 \fi
}%
\providecommand \natexlab [1]{#1}%
\providecommand \enquote  [1]{``#1''}%
\providecommand \bibnamefont  [1]{#1}%
\providecommand \bibfnamefont [1]{#1}%
\providecommand \citenamefont [1]{#1}%
\providecommand \href@noop [0]{\@secondoftwo}%
\providecommand \href [0]{\begingroup \@sanitize@url \@href}%
\providecommand \@href[1]{\@@startlink{#1}\@@href}%
\providecommand \@@href[1]{\endgroup#1\@@endlink}%
\providecommand \@sanitize@url [0]{\catcode `\\12\catcode `\$12\catcode `\&12\catcode `\#12\catcode `\^12\catcode `\_12\catcode `\%12\relax}%
\providecommand \@@startlink[1]{}%
\providecommand \@@endlink[0]{}%
\providecommand \url  [0]{\begingroup\@sanitize@url \@url }%
\providecommand \@url [1]{\endgroup\@href {#1}{\urlprefix }}%
\providecommand \urlprefix  [0]{URL }%
\providecommand \Eprint [0]{\href }%
\providecommand \doibase [0]{http://dx.doi.org/}%
\providecommand \selectlanguage [0]{\@gobble}%
\providecommand \bibinfo  [0]{\@secondoftwo}%
\providecommand \bibfield  [0]{\@secondoftwo}%
\providecommand \translation [1]{[#1]}%
\providecommand \BibitemOpen [0]{}%
\providecommand \bibitemStop [0]{}%
\providecommand \bibitemNoStop [0]{.\EOS\space}%
\providecommand \EOS [0]{\spacefactor3000\relax}%
\providecommand \BibitemShut  [1]{\csname bibitem#1\endcsname}%
\let\auto@bib@innerbib\@empty
\bibitem [{\citenamefont {Leone}\ \emph {et~al.}(2014)\citenamefont {Leone}, \citenamefont {McCurdy}, \citenamefont {Burgd{\"o}rfer}, \citenamefont {Cederbaum}, \citenamefont {Chang}, \citenamefont {Dudovich}, \citenamefont {Feist}, \citenamefont {Greene}, \citenamefont {Ivanov}, \citenamefont {Kienberger} \emph {et~al.}}]{leone2014will}%
  \BibitemOpen
  \bibfield  {author} {\bibinfo {author} {\bibfnamefont {Stephen~R}\ \bibnamefont {Leone}}, \bibinfo {author} {\bibfnamefont {C~William}\ \bibnamefont {McCurdy}}, \bibinfo {author} {\bibfnamefont {Joachim}\ \bibnamefont {Burgd{\"o}rfer}}, \bibinfo {author} {\bibfnamefont {Lorenz~S}\ \bibnamefont {Cederbaum}}, \bibinfo {author} {\bibfnamefont {Zenghu}\ \bibnamefont {Chang}}, \bibinfo {author} {\bibfnamefont {Nirit}\ \bibnamefont {Dudovich}}, \bibinfo {author} {\bibfnamefont {Johannes}\ \bibnamefont {Feist}}, \bibinfo {author} {\bibfnamefont {Chris~H}\ \bibnamefont {Greene}}, \bibinfo {author} {\bibfnamefont {Misha}\ \bibnamefont {Ivanov}}, \bibinfo {author} {\bibfnamefont {Reinhard}\ \bibnamefont {Kienberger}},  \emph {et~al.},\ }\bibfield  {title} {\enquote {\bibinfo {title} {What will it take to observe processes in'real time'?}}\ }\href@noop {} {\bibfield  {journal} {\bibinfo  {journal} {Nature Photonics}\ }\textbf {\bibinfo {volume} {8}},\ \bibinfo {pages} {162--166} (\bibinfo {year} {2014})}\BibitemShut
  {NoStop}%
\bibitem [{\citenamefont {Ramasesha}\ \emph {et~al.}(2016)\citenamefont {Ramasesha}, \citenamefont {Leone},\ and\ \citenamefont {Neumark}}]{ramasesha2016real}%
  \BibitemOpen
  \bibfield  {author} {\bibinfo {author} {\bibfnamefont {Krupa}\ \bibnamefont {Ramasesha}}, \bibinfo {author} {\bibfnamefont {Stephen~R}\ \bibnamefont {Leone}}, \ and\ \bibinfo {author} {\bibfnamefont {Daniel~M}\ \bibnamefont {Neumark}},\ }\bibfield  {title} {\enquote {\bibinfo {title} {Real-time probing of electron dynamics using attosecond time-resolved spectroscopy},}\ }\href@noop {} {\bibfield  {journal} {\bibinfo  {journal} {Annual Review of Physical Chemistry}\ }\textbf {\bibinfo {volume} {67}},\ \bibinfo {pages} {41--63} (\bibinfo {year} {2016})}\BibitemShut {NoStop}%
\bibitem [{\citenamefont {Liu}\ \emph {et~al.}(2017)\citenamefont {Liu}, \citenamefont {Li}, \citenamefont {You}, \citenamefont {Ghimire}, \citenamefont {Heinz},\ and\ \citenamefont {Reis}}]{liu2017high}%
  \BibitemOpen
  \bibfield  {author} {\bibinfo {author} {\bibfnamefont {Hanzhe}\ \bibnamefont {Liu}}, \bibinfo {author} {\bibfnamefont {Yilei}\ \bibnamefont {Li}}, \bibinfo {author} {\bibfnamefont {Yong~Sing}\ \bibnamefont {You}}, \bibinfo {author} {\bibfnamefont {Shambhu}\ \bibnamefont {Ghimire}}, \bibinfo {author} {\bibfnamefont {Tony~F}\ \bibnamefont {Heinz}}, \ and\ \bibinfo {author} {\bibfnamefont {David~A}\ \bibnamefont {Reis}},\ }\bibfield  {title} {\enquote {\bibinfo {title} {High-harmonic generation from an atomically thin semiconductor},}\ }\href@noop {} {\bibfield  {journal} {\bibinfo  {journal} {Nature Physics}\ }\textbf {\bibinfo {volume} {13}},\ \bibinfo {pages} {262--265} (\bibinfo {year} {2017})}\BibitemShut {NoStop}%
\bibitem [{\citenamefont {Ulbricht}\ \emph {et~al.}(2011)\citenamefont {Ulbricht}, \citenamefont {Hendry}, \citenamefont {Shan}, \citenamefont {Heinz},\ and\ \citenamefont {Bonn}}]{ulbricht2011carrier}%
  \BibitemOpen
  \bibfield  {author} {\bibinfo {author} {\bibfnamefont {Ronald}\ \bibnamefont {Ulbricht}}, \bibinfo {author} {\bibfnamefont {Euan}\ \bibnamefont {Hendry}}, \bibinfo {author} {\bibfnamefont {Jie}\ \bibnamefont {Shan}}, \bibinfo {author} {\bibfnamefont {Tony~F}\ \bibnamefont {Heinz}}, \ and\ \bibinfo {author} {\bibfnamefont {Mischa}\ \bibnamefont {Bonn}},\ }\bibfield  {title} {\enquote {\bibinfo {title} {Carrier dynamics in semiconductors studied with time-resolved terahertz spectroscopy},}\ }\href@noop {} {\bibfield  {journal} {\bibinfo  {journal} {Reviews of Modern Physics}\ }\textbf {\bibinfo {volume} {83}},\ \bibinfo {pages} {543--586} (\bibinfo {year} {2011})}\BibitemShut {NoStop}%
\bibitem [{\citenamefont {Boschini}\ \emph {et~al.}(2024)\citenamefont {Boschini}, \citenamefont {Zonno},\ and\ \citenamefont {Damascelli}}]{boschini2024time}%
  \BibitemOpen
  \bibfield  {author} {\bibinfo {author} {\bibfnamefont {Fabio}\ \bibnamefont {Boschini}}, \bibinfo {author} {\bibfnamefont {Marta}\ \bibnamefont {Zonno}}, \ and\ \bibinfo {author} {\bibfnamefont {Andrea}\ \bibnamefont {Damascelli}},\ }\bibfield  {title} {\enquote {\bibinfo {title} {Time-resolved arpes studies of quantum materials},}\ }\href@noop {} {\bibfield  {journal} {\bibinfo  {journal} {Reviews of Modern Physics}\ }\textbf {\bibinfo {volume} {96}},\ \bibinfo {pages} {015003} (\bibinfo {year} {2024})}\BibitemShut {NoStop}%
\bibitem [{\citenamefont {Cavalleri}(2018)}]{cavalleri2018photo}%
  \BibitemOpen
  \bibfield  {author} {\bibinfo {author} {\bibfnamefont {Andrea}\ \bibnamefont {Cavalleri}},\ }\bibfield  {title} {\enquote {\bibinfo {title} {Photo-induced superconductivity},}\ }\href@noop {} {\bibfield  {journal} {\bibinfo  {journal} {Contemporary Physics}\ }\textbf {\bibinfo {volume} {59}},\ \bibinfo {pages} {31--46} (\bibinfo {year} {2018})}\BibitemShut {NoStop}%
\bibitem [{\citenamefont {Chang~Lee}\ \emph {et~al.}(2024)\citenamefont {Chang~Lee}, \citenamefont {Yue}, \citenamefont {Gaarde}, \citenamefont {Chan},\ and\ \citenamefont {Qiu}}]{chang2024many}%
  \BibitemOpen
  \bibfield  {author} {\bibinfo {author} {\bibfnamefont {Victor}\ \bibnamefont {Chang~Lee}}, \bibinfo {author} {\bibfnamefont {Lun}\ \bibnamefont {Yue}}, \bibinfo {author} {\bibfnamefont {Mette~B}\ \bibnamefont {Gaarde}}, \bibinfo {author} {\bibfnamefont {Yang-hao}\ \bibnamefont {Chan}}, \ and\ \bibinfo {author} {\bibfnamefont {Diana~Y}\ \bibnamefont {Qiu}},\ }\bibfield  {title} {\enquote {\bibinfo {title} {Many-body enhancement of high-harmonic generation in monolayer mos2},}\ }\href@noop {} {\bibfield  {journal} {\bibinfo  {journal} {Nature Communications}\ }\textbf {\bibinfo {volume} {15}},\ \bibinfo {pages} {6228} (\bibinfo {year} {2024})}\BibitemShut {NoStop}%
\bibitem [{\citenamefont {Chan}\ \emph {et~al.}(2021{\natexlab{a}})\citenamefont {Chan}, \citenamefont {Qiu}, \citenamefont {da~Jornada},\ and\ \citenamefont {Louie}}]{chan2021giant}%
  \BibitemOpen
  \bibfield  {author} {\bibinfo {author} {\bibfnamefont {Yang-Hao}\ \bibnamefont {Chan}}, \bibinfo {author} {\bibfnamefont {Diana~Y}\ \bibnamefont {Qiu}}, \bibinfo {author} {\bibfnamefont {Felipe~H}\ \bibnamefont {da~Jornada}}, \ and\ \bibinfo {author} {\bibfnamefont {Steven~G}\ \bibnamefont {Louie}},\ }\bibfield  {title} {\enquote {\bibinfo {title} {Giant exciton-enhanced shift currents and direct current conduction with subbandgap photo excitations produced by many-electron interactions},}\ }\href@noop {} {\bibfield  {journal} {\bibinfo  {journal} {Proceedings of the National Academy of Sciences}\ }\textbf {\bibinfo {volume} {118}},\ \bibinfo {pages} {e1906938118} (\bibinfo {year} {2021}{\natexlab{a}})}\BibitemShut {NoStop}%
\bibitem [{\citenamefont {Chan}\ \emph {et~al.}(2023)\citenamefont {Chan}, \citenamefont {Qiu}, \citenamefont {da~Jornada},\ and\ \citenamefont {Louie}}]{chan2023giant}%
  \BibitemOpen
  \bibfield  {author} {\bibinfo {author} {\bibfnamefont {Y-H}\ \bibnamefont {Chan}}, \bibinfo {author} {\bibfnamefont {Diana~Y}\ \bibnamefont {Qiu}}, \bibinfo {author} {\bibfnamefont {Felipe~H}\ \bibnamefont {da~Jornada}}, \ and\ \bibinfo {author} {\bibfnamefont {Steven~G}\ \bibnamefont {Louie}},\ }\bibfield  {title} {\enquote {\bibinfo {title} {Giant self-driven exciton-floquet signatures in time-resolved photoemission spectroscopy of mos2 from time-dependent gw approach},}\ }\href@noop {} {\bibfield  {journal} {\bibinfo  {journal} {Proceedings of the National Academy of Sciences}\ }\textbf {\bibinfo {volume} {120}},\ \bibinfo {pages} {e2301957120} (\bibinfo {year} {2023})}\BibitemShut {NoStop}%
\bibitem [{\citenamefont {Man}\ \emph {et~al.}(2021)\citenamefont {Man}, \citenamefont {Mad{\'e}o}, \citenamefont {Sahoo}, \citenamefont {Xie}, \citenamefont {Campbell}, \citenamefont {Pareek}, \citenamefont {Karmakar}, \citenamefont {Wong}, \citenamefont {Al-Mahboob}, \citenamefont {Chan} \emph {et~al.}}]{man2021experimental}%
  \BibitemOpen
  \bibfield  {author} {\bibinfo {author} {\bibfnamefont {Michael~KL}\ \bibnamefont {Man}}, \bibinfo {author} {\bibfnamefont {Julien}\ \bibnamefont {Mad{\'e}o}}, \bibinfo {author} {\bibfnamefont {Chakradhar}\ \bibnamefont {Sahoo}}, \bibinfo {author} {\bibfnamefont {Kaichen}\ \bibnamefont {Xie}}, \bibinfo {author} {\bibfnamefont {Marshall}\ \bibnamefont {Campbell}}, \bibinfo {author} {\bibfnamefont {Vivek}\ \bibnamefont {Pareek}}, \bibinfo {author} {\bibfnamefont {Arka}\ \bibnamefont {Karmakar}}, \bibinfo {author} {\bibfnamefont {E~Laine}\ \bibnamefont {Wong}}, \bibinfo {author} {\bibfnamefont {Abdullah}\ \bibnamefont {Al-Mahboob}}, \bibinfo {author} {\bibfnamefont {Nicholas~S}\ \bibnamefont {Chan}},  \emph {et~al.},\ }\bibfield  {title} {\enquote {\bibinfo {title} {Experimental measurement of the intrinsic excitonic wave function},}\ }\href@noop {} {\bibfield  {journal} {\bibinfo  {journal} {Science Advances}\ }\textbf {\bibinfo {volume} {7}},\ \bibinfo {pages} {eabg0192} (\bibinfo {year} {2021})}\BibitemShut
  {NoStop}%
\bibitem [{\citenamefont {Mad{\'e}o}\ \emph {et~al.}(2020)\citenamefont {Mad{\'e}o}, \citenamefont {Man}, \citenamefont {Sahoo}, \citenamefont {Campbell}, \citenamefont {Pareek}, \citenamefont {Wong}, \citenamefont {Al-Mahboob}, \citenamefont {Chan}, \citenamefont {Karmakar}, \citenamefont {Mariserla} \emph {et~al.}}]{madeo2020directly}%
  \BibitemOpen
  \bibfield  {author} {\bibinfo {author} {\bibfnamefont {Julien}\ \bibnamefont {Mad{\'e}o}}, \bibinfo {author} {\bibfnamefont {Michael~KL}\ \bibnamefont {Man}}, \bibinfo {author} {\bibfnamefont {Chakradhar}\ \bibnamefont {Sahoo}}, \bibinfo {author} {\bibfnamefont {Marshall}\ \bibnamefont {Campbell}}, \bibinfo {author} {\bibfnamefont {Vivek}\ \bibnamefont {Pareek}}, \bibinfo {author} {\bibfnamefont {E~Laine}\ \bibnamefont {Wong}}, \bibinfo {author} {\bibfnamefont {Abdullah}\ \bibnamefont {Al-Mahboob}}, \bibinfo {author} {\bibfnamefont {Nicholas~S}\ \bibnamefont {Chan}}, \bibinfo {author} {\bibfnamefont {Arka}\ \bibnamefont {Karmakar}}, \bibinfo {author} {\bibfnamefont {Bala Murali~Krishna}\ \bibnamefont {Mariserla}},  \emph {et~al.},\ }\bibfield  {title} {\enquote {\bibinfo {title} {Directly visualizing the momentum-forbidden dark excitons and their dynamics in atomically thin semiconductors},}\ }\href@noop {} {\bibfield  {journal} {\bibinfo  {journal} {Science}\ }\textbf {\bibinfo {volume} {370}},\ \bibinfo
  {pages} {1199--1204} (\bibinfo {year} {2020})}\BibitemShut {NoStop}%
\bibitem [{\citenamefont {Lin}\ \emph {et~al.}(2022)\citenamefont {Lin}, \citenamefont {Chan}, \citenamefont {Lee}, \citenamefont {Lu}, \citenamefont {Li}, \citenamefont {Chang}, \citenamefont {Shih}, \citenamefont {Kaindl}, \citenamefont {Louie},\ and\ \citenamefont {Lanzara}}]{lin2022exciton}%
  \BibitemOpen
  \bibfield  {author} {\bibinfo {author} {\bibfnamefont {Yi}~\bibnamefont {Lin}}, \bibinfo {author} {\bibfnamefont {Yang-hao}\ \bibnamefont {Chan}}, \bibinfo {author} {\bibfnamefont {Woojoo}\ \bibnamefont {Lee}}, \bibinfo {author} {\bibfnamefont {Li-Syuan}\ \bibnamefont {Lu}}, \bibinfo {author} {\bibfnamefont {Zhenglu}\ \bibnamefont {Li}}, \bibinfo {author} {\bibfnamefont {Wen-Hao}\ \bibnamefont {Chang}}, \bibinfo {author} {\bibfnamefont {Chih-Kang}\ \bibnamefont {Shih}}, \bibinfo {author} {\bibfnamefont {Robert~A}\ \bibnamefont {Kaindl}}, \bibinfo {author} {\bibfnamefont {Steven~G}\ \bibnamefont {Louie}}, \ and\ \bibinfo {author} {\bibfnamefont {Alessandra}\ \bibnamefont {Lanzara}},\ }\bibfield  {title} {\enquote {\bibinfo {title} {Exciton-driven renormalization of quasiparticle band structure in monolayer mos 2},}\ }\href@noop {} {\bibfield  {journal} {\bibinfo  {journal} {Physical Review B}\ }\textbf {\bibinfo {volume} {106}},\ \bibinfo {pages} {L081117} (\bibinfo {year} {2022})}\BibitemShut {NoStop}%
\bibitem [{\citenamefont {Attaccalite}\ \emph {et~al.}(2011)\citenamefont {Attaccalite}, \citenamefont {Gr{\"u}ning},\ and\ \citenamefont {Marini}}]{attaccalite2011real}%
  \BibitemOpen
  \bibfield  {author} {\bibinfo {author} {\bibfnamefont {Claudio}\ \bibnamefont {Attaccalite}}, \bibinfo {author} {\bibfnamefont {M}~\bibnamefont {Gr{\"u}ning}}, \ and\ \bibinfo {author} {\bibfnamefont {A}~\bibnamefont {Marini}},\ }\bibfield  {title} {\enquote {\bibinfo {title} {Real-time approach to the optical properties of solids and nanostructures: Time-dependent bethe-salpeter equation},}\ }\href@noop {} {\bibfield  {journal} {\bibinfo  {journal} {Physical Review B—Condensed Matter and Materials Physics}\ }\textbf {\bibinfo {volume} {84}},\ \bibinfo {pages} {245110} (\bibinfo {year} {2011})}\BibitemShut {NoStop}%
\bibitem [{\citenamefont {Perfetto}\ \emph {et~al.}(2015)\citenamefont {Perfetto}, \citenamefont {Sangalli}, \citenamefont {Marini},\ and\ \citenamefont {Stefanucci}}]{perfetto2015nonequilibrium}%
  \BibitemOpen
  \bibfield  {author} {\bibinfo {author} {\bibfnamefont {E}~\bibnamefont {Perfetto}}, \bibinfo {author} {\bibfnamefont {D}~\bibnamefont {Sangalli}}, \bibinfo {author} {\bibfnamefont {A}~\bibnamefont {Marini}}, \ and\ \bibinfo {author} {\bibfnamefont {G}~\bibnamefont {Stefanucci}},\ }\bibfield  {title} {\enquote {\bibinfo {title} {Nonequilibrium bethe-salpeter equation for transient photoabsorption spectroscopy},}\ }\href@noop {} {\bibfield  {journal} {\bibinfo  {journal} {Physical Review B}\ }\textbf {\bibinfo {volume} {92}},\ \bibinfo {pages} {205304} (\bibinfo {year} {2015})}\BibitemShut {NoStop}%
\bibitem [{\citenamefont {Perfetto}\ \emph {et~al.}(2022)\citenamefont {Perfetto}, \citenamefont {Pavlyukh},\ and\ \citenamefont {Stefanucci}}]{perfetto2022real}%
  \BibitemOpen
  \bibfield  {author} {\bibinfo {author} {\bibfnamefont {Enrico}\ \bibnamefont {Perfetto}}, \bibinfo {author} {\bibfnamefont {Yaroslav}\ \bibnamefont {Pavlyukh}}, \ and\ \bibinfo {author} {\bibfnamefont {Gianluca}\ \bibnamefont {Stefanucci}},\ }\bibfield  {title} {\enquote {\bibinfo {title} {Real-time gw: Toward an ab initio description of the ultrafast carrier and exciton dynamics in two-dimensional materials},}\ }\href@noop {} {\bibfield  {journal} {\bibinfo  {journal} {Physical Review Letters}\ }\textbf {\bibinfo {volume} {128}},\ \bibinfo {pages} {016801} (\bibinfo {year} {2022})}\BibitemShut {NoStop}%
\bibitem [{\citenamefont {Marques}\ and\ \citenamefont {Gross}(2004)}]{marques2004time}%
  \BibitemOpen
  \bibfield  {author} {\bibinfo {author} {\bibfnamefont {Miguel~AL}\ \bibnamefont {Marques}}\ and\ \bibinfo {author} {\bibfnamefont {Eberhard~KU}\ \bibnamefont {Gross}},\ }\bibfield  {title} {\enquote {\bibinfo {title} {Time-dependent density functional theory},}\ }\href@noop {} {\bibfield  {journal} {\bibinfo  {journal} {Annu. Rev. Phys. Chem.}\ }\textbf {\bibinfo {volume} {55}},\ \bibinfo {pages} {427--455} (\bibinfo {year} {2004})}\BibitemShut {NoStop}%
\bibitem [{\citenamefont {Reining}\ \emph {et~al.}(2002)\citenamefont {Reining}, \citenamefont {Olevano}, \citenamefont {Rubio},\ and\ \citenamefont {Onida}}]{reining2002excitonic}%
  \BibitemOpen
  \bibfield  {author} {\bibinfo {author} {\bibfnamefont {Lucia}\ \bibnamefont {Reining}}, \bibinfo {author} {\bibfnamefont {Valerio}\ \bibnamefont {Olevano}}, \bibinfo {author} {\bibfnamefont {Angel}\ \bibnamefont {Rubio}}, \ and\ \bibinfo {author} {\bibfnamefont {Giovanni}\ \bibnamefont {Onida}},\ }\bibfield  {title} {\enquote {\bibinfo {title} {Excitonic effects in solids described by time-dependent density-functional theory},}\ }\href@noop {} {\bibfield  {journal} {\bibinfo  {journal} {Physical review letters}\ }\textbf {\bibinfo {volume} {88}},\ \bibinfo {pages} {066404} (\bibinfo {year} {2002})}\BibitemShut {NoStop}%
\bibitem [{\citenamefont {Romaniello}\ \emph {et~al.}(2009)\citenamefont {Romaniello}, \citenamefont {Sangalli}, \citenamefont {Berger}, \citenamefont {Sottile}, \citenamefont {Molinari}, \citenamefont {Reining},\ and\ \citenamefont {Onida}}]{romaniello2009double}%
  \BibitemOpen
  \bibfield  {author} {\bibinfo {author} {\bibfnamefont {Pina}\ \bibnamefont {Romaniello}}, \bibinfo {author} {\bibfnamefont {Davide}\ \bibnamefont {Sangalli}}, \bibinfo {author} {\bibfnamefont {JA}~\bibnamefont {Berger}}, \bibinfo {author} {\bibfnamefont {Francesco}\ \bibnamefont {Sottile}}, \bibinfo {author} {\bibfnamefont {Luca~G}\ \bibnamefont {Molinari}}, \bibinfo {author} {\bibfnamefont {Lucia}\ \bibnamefont {Reining}}, \ and\ \bibinfo {author} {\bibfnamefont {Giovanni}\ \bibnamefont {Onida}},\ }\bibfield  {title} {\enquote {\bibinfo {title} {Double excitations in finite systems},}\ }\href@noop {} {\bibfield  {journal} {\bibinfo  {journal} {The Journal of Chemical Physics}\ }\textbf {\bibinfo {volume} {130}} (\bibinfo {year} {2009})}\BibitemShut {NoStop}%
\bibitem [{\citenamefont {Sangalli}(2021)}]{sangalli2021excitons}%
  \BibitemOpen
  \bibfield  {author} {\bibinfo {author} {\bibfnamefont {D}~\bibnamefont {Sangalli}},\ }\bibfield  {title} {\enquote {\bibinfo {title} {Excitons and carriers in transient absorption and time-resolved arpes spectroscopy: An ab initio approach},}\ }\href@noop {} {\bibfield  {journal} {\bibinfo  {journal} {Physical Review Materials}\ }\textbf {\bibinfo {volume} {5}},\ \bibinfo {pages} {083803} (\bibinfo {year} {2021})}\BibitemShut {NoStop}%
\bibitem [{\citenamefont {Luo}\ \emph {et~al.}(2024)\citenamefont {Luo}, \citenamefont {Desai}, \citenamefont {Chang}, \citenamefont {Park},\ and\ \citenamefont {Bernardi}}]{luo2024data}%
  \BibitemOpen
  \bibfield  {author} {\bibinfo {author} {\bibfnamefont {Yao}\ \bibnamefont {Luo}}, \bibinfo {author} {\bibfnamefont {Dhruv}\ \bibnamefont {Desai}}, \bibinfo {author} {\bibfnamefont {Benjamin~K}\ \bibnamefont {Chang}}, \bibinfo {author} {\bibfnamefont {Jinsoo}\ \bibnamefont {Park}}, \ and\ \bibinfo {author} {\bibfnamefont {Marco}\ \bibnamefont {Bernardi}},\ }\bibfield  {title} {\enquote {\bibinfo {title} {Data-driven compression of electron-phonon interactions},}\ }\href@noop {} {\bibfield  {journal} {\bibinfo  {journal} {Physical Review X}\ }\textbf {\bibinfo {volume} {14}},\ \bibinfo {pages} {021023} (\bibinfo {year} {2024})}\BibitemShut {NoStop}%
\bibitem [{\citenamefont {Lu}\ and\ \citenamefont {Ying}(2015)}]{lu2015compression}%
  \BibitemOpen
  \bibfield  {author} {\bibinfo {author} {\bibfnamefont {Jianfeng}\ \bibnamefont {Lu}}\ and\ \bibinfo {author} {\bibfnamefont {Lexing}\ \bibnamefont {Ying}},\ }\bibfield  {title} {\enquote {\bibinfo {title} {Compression of the electron repulsion integral tensor in tensor hypercontraction format with cubic scaling cost},}\ }\href@noop {} {\bibfield  {journal} {\bibinfo  {journal} {Journal of Computational Physics}\ }\textbf {\bibinfo {volume} {302}},\ \bibinfo {pages} {329--335} (\bibinfo {year} {2015})}\BibitemShut {NoStop}%
\bibitem [{\citenamefont {Shao}\ \emph {et~al.}(2016)\citenamefont {Shao}, \citenamefont {Lin}, \citenamefont {Yang}, \citenamefont {Liu}, \citenamefont {Da~Jornada}, \citenamefont {Deslippe},\ and\ \citenamefont {Louie}}]{shao2016low}%
  \BibitemOpen
  \bibfield  {author} {\bibinfo {author} {\bibfnamefont {MeiYue}\ \bibnamefont {Shao}}, \bibinfo {author} {\bibfnamefont {Lin}\ \bibnamefont {Lin}}, \bibinfo {author} {\bibfnamefont {Chao}\ \bibnamefont {Yang}}, \bibinfo {author} {\bibfnamefont {Fang}\ \bibnamefont {Liu}}, \bibinfo {author} {\bibfnamefont {Felipe~H}\ \bibnamefont {Da~Jornada}}, \bibinfo {author} {\bibfnamefont {Jack}\ \bibnamefont {Deslippe}}, \ and\ \bibinfo {author} {\bibfnamefont {Steven~G}\ \bibnamefont {Louie}},\ }\bibfield  {title} {\enquote {\bibinfo {title} {Low rank approximation in g 0 w 0 calculations},}\ }\href@noop {} {\bibfield  {journal} {\bibinfo  {journal} {Science China Mathematics}\ }\textbf {\bibinfo {volume} {59}},\ \bibinfo {pages} {1593--1612} (\bibinfo {year} {2016})}\BibitemShut {NoStop}%
\bibitem [{\citenamefont {Kaye}\ \emph {et~al.}(2022)\citenamefont {Kaye}, \citenamefont {Chen},\ and\ \citenamefont {Parcollet}}]{Kaye2022GreenFunction}%
  \BibitemOpen
  \bibfield  {author} {\bibinfo {author} {\bibfnamefont {Jason}\ \bibnamefont {Kaye}}, \bibinfo {author} {\bibfnamefont {Kun}\ \bibnamefont {Chen}}, \ and\ \bibinfo {author} {\bibfnamefont {Olivier}\ \bibnamefont {Parcollet}},\ }\bibfield  {title} {\enquote {\bibinfo {title} {Discrete lehmann representation of imaginary time green's functions},}\ }\href {\doibase 10.1103/PhysRevB.105.235115} {\bibfield  {journal} {\bibinfo  {journal} {Phys. Rev. B}\ }\textbf {\bibinfo {volume} {105}},\ \bibinfo {pages} {235115} (\bibinfo {year} {2022})}\BibitemShut {NoStop}%
\bibitem [{\citenamefont {Dunlap}(2000)}]{dunlap2000robust}%
  \BibitemOpen
  \bibfield  {author} {\bibinfo {author} {\bibfnamefont {Brett~I}\ \bibnamefont {Dunlap}},\ }\bibfield  {title} {\enquote {\bibinfo {title} {Robust and variational fitting},}\ }\href@noop {} {\bibfield  {journal} {\bibinfo  {journal} {Physical Chemistry Chemical Physics}\ }\textbf {\bibinfo {volume} {2}},\ \bibinfo {pages} {2113--2116} (\bibinfo {year} {2000})}\BibitemShut {NoStop}%
\bibitem [{\citenamefont {Pham}\ and\ \citenamefont {Gordon}(2019)}]{pham2019compressing}%
  \BibitemOpen
  \bibfield  {author} {\bibinfo {author} {\bibfnamefont {Buu~Q}\ \bibnamefont {Pham}}\ and\ \bibinfo {author} {\bibfnamefont {Mark~S}\ \bibnamefont {Gordon}},\ }\bibfield  {title} {\enquote {\bibinfo {title} {Compressing the four-index two-electron repulsion integral matrix using the resolution-of-the-identity approximation combined with the rank factorization approximation},}\ }\href@noop {} {\bibfield  {journal} {\bibinfo  {journal} {Journal of Chemical Theory and Computation}\ }\textbf {\bibinfo {volume} {15}},\ \bibinfo {pages} {2254--2264} (\bibinfo {year} {2019})}\BibitemShut {NoStop}%
\bibitem [{\citenamefont {Del~Ben}\ \emph {et~al.}(2019)\citenamefont {Del~Ben}, \citenamefont {da~Jornada}, \citenamefont {Antonius}, \citenamefont {Rangel}, \citenamefont {Louie}, \citenamefont {Deslippe},\ and\ \citenamefont {Canning}}]{del2019static}%
  \BibitemOpen
  \bibfield  {author} {\bibinfo {author} {\bibfnamefont {Mauro}\ \bibnamefont {Del~Ben}}, \bibinfo {author} {\bibfnamefont {Felipe~H}\ \bibnamefont {da~Jornada}}, \bibinfo {author} {\bibfnamefont {Gabriel}\ \bibnamefont {Antonius}}, \bibinfo {author} {\bibfnamefont {Tonatiuh}\ \bibnamefont {Rangel}}, \bibinfo {author} {\bibfnamefont {Steven~G}\ \bibnamefont {Louie}}, \bibinfo {author} {\bibfnamefont {Jack}\ \bibnamefont {Deslippe}}, \ and\ \bibinfo {author} {\bibfnamefont {Andrew}\ \bibnamefont {Canning}},\ }\bibfield  {title} {\enquote {\bibinfo {title} {Static subspace approximation for the evaluation of g 0 w 0 quasiparticle energies within a sum-over-bands approach},}\ }\href@noop {} {\bibfield  {journal} {\bibinfo  {journal} {Physical Review B}\ }\textbf {\bibinfo {volume} {99}},\ \bibinfo {pages} {125128} (\bibinfo {year} {2019})}\BibitemShut {NoStop}%
\bibitem [{\citenamefont {Zang}\ \emph {et~al.}(2024)\citenamefont {Zang}, \citenamefont {Medvidovi{\'c}}, \citenamefont {Kiese}, \citenamefont {Di~Sante}, \citenamefont {Sengupta},\ and\ \citenamefont {Millis}}]{zang2024machine}%
  \BibitemOpen
  \bibfield  {author} {\bibinfo {author} {\bibfnamefont {Jiawei}\ \bibnamefont {Zang}}, \bibinfo {author} {\bibfnamefont {Matija}\ \bibnamefont {Medvidovi{\'c}}}, \bibinfo {author} {\bibfnamefont {Dominik}\ \bibnamefont {Kiese}}, \bibinfo {author} {\bibfnamefont {Domenico}\ \bibnamefont {Di~Sante}}, \bibinfo {author} {\bibfnamefont {Anirvan~M}\ \bibnamefont {Sengupta}}, \ and\ \bibinfo {author} {\bibfnamefont {Andrew~J}\ \bibnamefont {Millis}},\ }\bibfield  {title} {\enquote {\bibinfo {title} {Machine learning-based compression of quantum many body physics: Pca and autoencoder representation of the vertex function},}\ }\href@noop {} {\bibfield  {journal} {\bibinfo  {journal} {arXiv preprint arXiv:2403.15372}\ } (\bibinfo {year} {2024})}\BibitemShut {NoStop}%
\bibitem [{\citenamefont {Shinaoka}\ \emph {et~al.}(2017)\citenamefont {Shinaoka}, \citenamefont {Otsuki}, \citenamefont {Ohzeki},\ and\ \citenamefont {Yoshimi}}]{Shinaoka2017}%
  \BibitemOpen
  \bibfield  {author} {\bibinfo {author} {\bibfnamefont {Hiroshi}\ \bibnamefont {Shinaoka}}, \bibinfo {author} {\bibfnamefont {Junya}\ \bibnamefont {Otsuki}}, \bibinfo {author} {\bibfnamefont {Masayuki}\ \bibnamefont {Ohzeki}}, \ and\ \bibinfo {author} {\bibfnamefont {Kazuyoshi}\ \bibnamefont {Yoshimi}},\ }\bibfield  {title} {\enquote {\bibinfo {title} {Compressing green's function using intermediate representation between imaginary-time and real-frequency domains},}\ }\href {\doibase 10.1103/PhysRevB.96.035147} {\bibfield  {journal} {\bibinfo  {journal} {Phys. Rev. B}\ }\textbf {\bibinfo {volume} {96}},\ \bibinfo {pages} {035147} (\bibinfo {year} {2017})}\BibitemShut {NoStop}%
\bibitem [{\citenamefont {Zadoks}\ \emph {et~al.}(2024)\citenamefont {Zadoks}, \citenamefont {Marrazzo},\ and\ \citenamefont {Marzari}}]{zadoks2024spectral}%
  \BibitemOpen
  \bibfield  {author} {\bibinfo {author} {\bibfnamefont {Austin}\ \bibnamefont {Zadoks}}, \bibinfo {author} {\bibfnamefont {Antimo}\ \bibnamefont {Marrazzo}}, \ and\ \bibinfo {author} {\bibfnamefont {Nicola}\ \bibnamefont {Marzari}},\ }\bibfield  {title} {\enquote {\bibinfo {title} {Spectral operator representations},}\ }\href@noop {} {\bibfield  {journal} {\bibinfo  {journal} {arXiv preprint arXiv:2403.01514}\ } (\bibinfo {year} {2024})}\BibitemShut {NoStop}%
\bibitem [{\citenamefont {Kn{\o}sgaard}\ and\ \citenamefont {Thygesen}(2022)}]{knosgaard2022representing}%
  \BibitemOpen
  \bibfield  {author} {\bibinfo {author} {\bibfnamefont {Nikolaj~R{\o}rb{\ae}k}\ \bibnamefont {Kn{\o}sgaard}}\ and\ \bibinfo {author} {\bibfnamefont {Kristian~Sommer}\ \bibnamefont {Thygesen}},\ }\bibfield  {title} {\enquote {\bibinfo {title} {Representing individual electronic states for machine learning gw band structures of 2d materials},}\ }\href@noop {} {\bibfield  {journal} {\bibinfo  {journal} {Nature Communications}\ }\textbf {\bibinfo {volume} {13}},\ \bibinfo {pages} {468} (\bibinfo {year} {2022})}\BibitemShut {NoStop}%
\bibitem [{\citenamefont {Hou}\ \emph {et~al.}(2024)\citenamefont {Hou}, \citenamefont {Wu},\ and\ \citenamefont {Qiu}}]{hou2024unsupervised}%
  \BibitemOpen
  \bibfield  {author} {\bibinfo {author} {\bibfnamefont {Bowen}\ \bibnamefont {Hou}}, \bibinfo {author} {\bibfnamefont {Jinyuan}\ \bibnamefont {Wu}}, \ and\ \bibinfo {author} {\bibfnamefont {Diana~Y}\ \bibnamefont {Qiu}},\ }\bibfield  {title} {\enquote {\bibinfo {title} {Unsupervised representation learning of kohn--sham states and consequences for downstream predictions of many-body effects},}\ }\href@noop {} {\bibfield  {journal} {\bibinfo  {journal} {Nature Communications}\ }\textbf {\bibinfo {volume} {15}},\ \bibinfo {pages} {9481} (\bibinfo {year} {2024})}\BibitemShut {NoStop}%
\bibitem [{\citenamefont {Luchnikov}\ \emph {et~al.}(2019)\citenamefont {Luchnikov}, \citenamefont {Ryzhov}, \citenamefont {Stas}, \citenamefont {Filippov},\ and\ \citenamefont {Ouerdane}}]{luchnikov2019variational}%
  \BibitemOpen
  \bibfield  {author} {\bibinfo {author} {\bibfnamefont {Ilia~A}\ \bibnamefont {Luchnikov}}, \bibinfo {author} {\bibfnamefont {Alexander}\ \bibnamefont {Ryzhov}}, \bibinfo {author} {\bibfnamefont {Pieter-Jan}\ \bibnamefont {Stas}}, \bibinfo {author} {\bibfnamefont {Sergey~N}\ \bibnamefont {Filippov}}, \ and\ \bibinfo {author} {\bibfnamefont {Henni}\ \bibnamefont {Ouerdane}},\ }\bibfield  {title} {\enquote {\bibinfo {title} {Variational autoencoder reconstruction of complex many-body physics},}\ }\href@noop {} {\bibfield  {journal} {\bibinfo  {journal} {Entropy}\ }\textbf {\bibinfo {volume} {21}},\ \bibinfo {pages} {1091} (\bibinfo {year} {2019})}\BibitemShut {NoStop}%
\bibitem [{\citenamefont {Batzner}\ \emph {et~al.}(2022)\citenamefont {Batzner}, \citenamefont {Musaelian}, \citenamefont {Sun}, \citenamefont {Geiger}, \citenamefont {Mailoa}, \citenamefont {Kornbluth}, \citenamefont {Molinari}, \citenamefont {Smidt},\ and\ \citenamefont {Kozinsky}}]{batzner20223}%
  \BibitemOpen
  \bibfield  {author} {\bibinfo {author} {\bibfnamefont {Simon}\ \bibnamefont {Batzner}}, \bibinfo {author} {\bibfnamefont {Albert}\ \bibnamefont {Musaelian}}, \bibinfo {author} {\bibfnamefont {Lixin}\ \bibnamefont {Sun}}, \bibinfo {author} {\bibfnamefont {Mario}\ \bibnamefont {Geiger}}, \bibinfo {author} {\bibfnamefont {Jonathan~P}\ \bibnamefont {Mailoa}}, \bibinfo {author} {\bibfnamefont {Mordechai}\ \bibnamefont {Kornbluth}}, \bibinfo {author} {\bibfnamefont {Nicola}\ \bibnamefont {Molinari}}, \bibinfo {author} {\bibfnamefont {Tess~E}\ \bibnamefont {Smidt}}, \ and\ \bibinfo {author} {\bibfnamefont {Boris}\ \bibnamefont {Kozinsky}},\ }\bibfield  {title} {\enquote {\bibinfo {title} {E (3)-equivariant graph neural networks for data-efficient and accurate interatomic potentials},}\ }\href@noop {} {\bibfield  {journal} {\bibinfo  {journal} {Nature communications}\ }\textbf {\bibinfo {volume} {13}},\ \bibinfo {pages} {2453} (\bibinfo {year} {2022})}\BibitemShut {NoStop}%
\bibitem [{\citenamefont {Thomas}\ \emph {et~al.}(2018)\citenamefont {Thomas}, \citenamefont {Smidt}, \citenamefont {Kearnes}, \citenamefont {Yang}, \citenamefont {Li}, \citenamefont {Kohlhoff},\ and\ \citenamefont {Riley}}]{thomas2018tensor}%
  \BibitemOpen
  \bibfield  {author} {\bibinfo {author} {\bibfnamefont {Nathaniel}\ \bibnamefont {Thomas}}, \bibinfo {author} {\bibfnamefont {Tess}\ \bibnamefont {Smidt}}, \bibinfo {author} {\bibfnamefont {Steven}\ \bibnamefont {Kearnes}}, \bibinfo {author} {\bibfnamefont {Lusann}\ \bibnamefont {Yang}}, \bibinfo {author} {\bibfnamefont {Li}~\bibnamefont {Li}}, \bibinfo {author} {\bibfnamefont {Kai}\ \bibnamefont {Kohlhoff}}, \ and\ \bibinfo {author} {\bibfnamefont {Patrick}\ \bibnamefont {Riley}},\ }\bibfield  {title} {\enquote {\bibinfo {title} {Tensor field networks: Rotation-and translation-equivariant neural networks for 3d point clouds},}\ }\href@noop {} {\bibfield  {journal} {\bibinfo  {journal} {arXiv preprint arXiv:1802.08219}\ } (\bibinfo {year} {2018})}\BibitemShut {NoStop}%
\bibitem [{\citenamefont {Li}\ \emph {et~al.}(2024)\citenamefont {Li}, \citenamefont {Tang}, \citenamefont {Fu}, \citenamefont {Dong}, \citenamefont {Zou}, \citenamefont {Gong}, \citenamefont {Duan},\ and\ \citenamefont {Xu}}]{li2024deep}%
  \BibitemOpen
  \bibfield  {author} {\bibinfo {author} {\bibfnamefont {He}~\bibnamefont {Li}}, \bibinfo {author} {\bibfnamefont {Zechen}\ \bibnamefont {Tang}}, \bibinfo {author} {\bibfnamefont {Jingheng}\ \bibnamefont {Fu}}, \bibinfo {author} {\bibfnamefont {Wen-Han}\ \bibnamefont {Dong}}, \bibinfo {author} {\bibfnamefont {Nianlong}\ \bibnamefont {Zou}}, \bibinfo {author} {\bibfnamefont {Xiaoxun}\ \bibnamefont {Gong}}, \bibinfo {author} {\bibfnamefont {Wenhui}\ \bibnamefont {Duan}}, \ and\ \bibinfo {author} {\bibfnamefont {Yong}\ \bibnamefont {Xu}},\ }\bibfield  {title} {\enquote {\bibinfo {title} {Deep-learning density functional perturbation theory},}\ }\href@noop {} {\bibfield  {journal} {\bibinfo  {journal} {Physical Review Letters}\ }\textbf {\bibinfo {volume} {132}},\ \bibinfo {pages} {096401} (\bibinfo {year} {2024})}\BibitemShut {NoStop}%
\bibitem [{\citenamefont {Yang}\ \emph {et~al.}(2020)\citenamefont {Yang}, \citenamefont {Leng}, \citenamefont {Yu}, \citenamefont {Patel}, \citenamefont {Hu},\ and\ \citenamefont {Pu}}]{yang2020deep}%
  \BibitemOpen
  \bibfield  {author} {\bibinfo {author} {\bibfnamefont {Li}~\bibnamefont {Yang}}, \bibinfo {author} {\bibfnamefont {Zhaoqi}\ \bibnamefont {Leng}}, \bibinfo {author} {\bibfnamefont {Guangyuan}\ \bibnamefont {Yu}}, \bibinfo {author} {\bibfnamefont {Ankit}\ \bibnamefont {Patel}}, \bibinfo {author} {\bibfnamefont {Wen-Jun}\ \bibnamefont {Hu}}, \ and\ \bibinfo {author} {\bibfnamefont {Han}\ \bibnamefont {Pu}},\ }\bibfield  {title} {\enquote {\bibinfo {title} {Deep learning-enhanced variational monte carlo method for quantum many-body physics},}\ }\href@noop {} {\bibfield  {journal} {\bibinfo  {journal} {Physical Review Research}\ }\textbf {\bibinfo {volume} {2}},\ \bibinfo {pages} {012039} (\bibinfo {year} {2020})}\BibitemShut {NoStop}%
\bibitem [{\citenamefont {Medvidovi{\'c}}\ and\ \citenamefont {Moreno}(2024)}]{medvidovic2024neural}%
  \BibitemOpen
  \bibfield  {author} {\bibinfo {author} {\bibfnamefont {Matija}\ \bibnamefont {Medvidovi{\'c}}}\ and\ \bibinfo {author} {\bibfnamefont {Javier~Robledo}\ \bibnamefont {Moreno}},\ }\bibfield  {title} {\enquote {\bibinfo {title} {Neural-network quantum states for many-body physics},}\ }\href@noop {} {\bibfield  {journal} {\bibinfo  {journal} {arXiv preprint arXiv:2402.11014}\ } (\bibinfo {year} {2024})}\BibitemShut {NoStop}%
\bibitem [{\citenamefont {No{\'e}}\ \emph {et~al.}(2019)\citenamefont {No{\'e}}, \citenamefont {Olsson}, \citenamefont {K{\"o}hler},\ and\ \citenamefont {Wu}}]{noe2019boltzmann}%
  \BibitemOpen
  \bibfield  {author} {\bibinfo {author} {\bibfnamefont {Frank}\ \bibnamefont {No{\'e}}}, \bibinfo {author} {\bibfnamefont {Simon}\ \bibnamefont {Olsson}}, \bibinfo {author} {\bibfnamefont {Jonas}\ \bibnamefont {K{\"o}hler}}, \ and\ \bibinfo {author} {\bibfnamefont {Hao}\ \bibnamefont {Wu}},\ }\bibfield  {title} {\enquote {\bibinfo {title} {Boltzmann generators: Sampling equilibrium states of many-body systems with deep learning},}\ }\href@noop {} {\bibfield  {journal} {\bibinfo  {journal} {Science}\ }\textbf {\bibinfo {volume} {365}},\ \bibinfo {pages} {eaaw1147} (\bibinfo {year} {2019})}\BibitemShut {NoStop}%
\bibitem [{\citenamefont {Carrasquilla}\ and\ \citenamefont {Torlai}(2021)}]{carrasquilla2021use}%
  \BibitemOpen
  \bibfield  {author} {\bibinfo {author} {\bibfnamefont {Juan}\ \bibnamefont {Carrasquilla}}\ and\ \bibinfo {author} {\bibfnamefont {Giacomo}\ \bibnamefont {Torlai}},\ }\bibfield  {title} {\enquote {\bibinfo {title} {How to use neural networks to investigate quantum many-body physics},}\ }\href@noop {} {\bibfield  {journal} {\bibinfo  {journal} {PRX Quantum}\ }\textbf {\bibinfo {volume} {2}},\ \bibinfo {pages} {040201} (\bibinfo {year} {2021})}\BibitemShut {NoStop}%
\bibitem [{\citenamefont {Yin}\ \emph {et~al.}(2023)\citenamefont {Yin}, \citenamefont {Chan}, \citenamefont {da~Jornada}, \citenamefont {Qiu}, \citenamefont {Yang},\ and\ \citenamefont {Louie}}]{yin2023analyzing}%
  \BibitemOpen
  \bibfield  {author} {\bibinfo {author} {\bibfnamefont {Jia}\ \bibnamefont {Yin}}, \bibinfo {author} {\bibfnamefont {Yang-hao}\ \bibnamefont {Chan}}, \bibinfo {author} {\bibfnamefont {Felipe~H}\ \bibnamefont {da~Jornada}}, \bibinfo {author} {\bibfnamefont {Diana~Y}\ \bibnamefont {Qiu}}, \bibinfo {author} {\bibfnamefont {Chao}\ \bibnamefont {Yang}}, \ and\ \bibinfo {author} {\bibfnamefont {Steven~G}\ \bibnamefont {Louie}},\ }\bibfield  {title} {\enquote {\bibinfo {title} {Analyzing and predicting non-equilibrium many-body dynamics via dynamic mode decomposition},}\ }\href@noop {} {\bibfield  {journal} {\bibinfo  {journal} {Journal of Computational Physics}\ }\textbf {\bibinfo {volume} {477}},\ \bibinfo {pages} {111909} (\bibinfo {year} {2023})}\BibitemShut {NoStop}%
\bibitem [{\citenamefont {Maliyov}\ \emph {et~al.}(2024)\citenamefont {Maliyov}, \citenamefont {Yin}, \citenamefont {Yao}, \citenamefont {Yang},\ and\ \citenamefont {Bernardi}}]{maliyov2024dynamic}%
  \BibitemOpen
  \bibfield  {author} {\bibinfo {author} {\bibfnamefont {Ivan}\ \bibnamefont {Maliyov}}, \bibinfo {author} {\bibfnamefont {Jia}\ \bibnamefont {Yin}}, \bibinfo {author} {\bibfnamefont {Jia}\ \bibnamefont {Yao}}, \bibinfo {author} {\bibfnamefont {Chao}\ \bibnamefont {Yang}}, \ and\ \bibinfo {author} {\bibfnamefont {Marco}\ \bibnamefont {Bernardi}},\ }\bibfield  {title} {\enquote {\bibinfo {title} {Dynamic mode decomposition of nonequilibrium electron-phonon dynamics: accelerating the first-principles real-time boltzmann equation},}\ }\href@noop {} {\bibfield  {journal} {\bibinfo  {journal} {npj Computational Materials}\ }\textbf {\bibinfo {volume} {10}},\ \bibinfo {pages} {123} (\bibinfo {year} {2024})}\BibitemShut {NoStop}%
\bibitem [{\citenamefont {Reeves}\ \emph {et~al.}(2023)\citenamefont {Reeves}, \citenamefont {Yin}, \citenamefont {Zhu}, \citenamefont {Ibrahim}, \citenamefont {Yang},\ and\ \citenamefont {Vl{\v{c}}ek}}]{reeves2023dynamic}%
  \BibitemOpen
  \bibfield  {author} {\bibinfo {author} {\bibfnamefont {Cian~C}\ \bibnamefont {Reeves}}, \bibinfo {author} {\bibfnamefont {Jia}\ \bibnamefont {Yin}}, \bibinfo {author} {\bibfnamefont {Yuanran}\ \bibnamefont {Zhu}}, \bibinfo {author} {\bibfnamefont {Khaled~Z}\ \bibnamefont {Ibrahim}}, \bibinfo {author} {\bibfnamefont {Chao}\ \bibnamefont {Yang}}, \ and\ \bibinfo {author} {\bibfnamefont {Vojt{\v{e}}ch}\ \bibnamefont {Vl{\v{c}}ek}},\ }\bibfield  {title} {\enquote {\bibinfo {title} {Dynamic mode decomposition for extrapolating nonequilibrium green's-function dynamics},}\ }\href@noop {} {\bibfield  {journal} {\bibinfo  {journal} {Physical Review B}\ }\textbf {\bibinfo {volume} {107}},\ \bibinfo {pages} {075107} (\bibinfo {year} {2023})}\BibitemShut {NoStop}%
\bibitem [{\citenamefont {Gu}\ \emph {et~al.}(2024)\citenamefont {Gu}, \citenamefont {Lin}, \citenamefont {Lee},\ and\ \citenamefont {Qiu}}]{gu2024probabilistic}%
  \BibitemOpen
  \bibfield  {author} {\bibinfo {author} {\bibfnamefont {Mengyang}\ \bibnamefont {Gu}}, \bibinfo {author} {\bibfnamefont {Yizi}\ \bibnamefont {Lin}}, \bibinfo {author} {\bibfnamefont {Victor~Chang}\ \bibnamefont {Lee}}, \ and\ \bibinfo {author} {\bibfnamefont {Diana~Y}\ \bibnamefont {Qiu}},\ }\bibfield  {title} {\enquote {\bibinfo {title} {Probabilistic forecast of nonlinear dynamical systems with uncertainty quantification},}\ }\href@noop {} {\bibfield  {journal} {\bibinfo  {journal} {Physica D: Nonlinear Phenomena}\ }\textbf {\bibinfo {volume} {457}},\ \bibinfo {pages} {133938} (\bibinfo {year} {2024})}\BibitemShut {NoStop}%
\bibitem [{\citenamefont {Lu}\ \emph {et~al.}(2021)\citenamefont {Lu}, \citenamefont {Jin}, \citenamefont {Pang}, \citenamefont {Zhang},\ and\ \citenamefont {Karniadakis}}]{lu2021learning}%
  \BibitemOpen
  \bibfield  {author} {\bibinfo {author} {\bibfnamefont {Lu}~\bibnamefont {Lu}}, \bibinfo {author} {\bibfnamefont {Pengzhan}\ \bibnamefont {Jin}}, \bibinfo {author} {\bibfnamefont {Guofei}\ \bibnamefont {Pang}}, \bibinfo {author} {\bibfnamefont {Zhongqiang}\ \bibnamefont {Zhang}}, \ and\ \bibinfo {author} {\bibfnamefont {George~Em}\ \bibnamefont {Karniadakis}},\ }\bibfield  {title} {\enquote {\bibinfo {title} {Learning nonlinear operators via deeponet based on the universal approximation theorem of operators},}\ }\href@noop {} {\bibfield  {journal} {\bibinfo  {journal} {Nature machine intelligence}\ }\textbf {\bibinfo {volume} {3}},\ \bibinfo {pages} {218--229} (\bibinfo {year} {2021})}\BibitemShut {NoStop}%
\bibitem [{\citenamefont {Li}\ \emph {et~al.}(2020{\natexlab{a}})\citenamefont {Li}, \citenamefont {Kovachki}, \citenamefont {Azizzadenesheli}, \citenamefont {Liu}, \citenamefont {Bhattacharya}, \citenamefont {Stuart},\ and\ \citenamefont {Anandkumar}}]{li2020fourier}%
  \BibitemOpen
  \bibfield  {author} {\bibinfo {author} {\bibfnamefont {Zongyi}\ \bibnamefont {Li}}, \bibinfo {author} {\bibfnamefont {Nikola}\ \bibnamefont {Kovachki}}, \bibinfo {author} {\bibfnamefont {Kamyar}\ \bibnamefont {Azizzadenesheli}}, \bibinfo {author} {\bibfnamefont {Burigede}\ \bibnamefont {Liu}}, \bibinfo {author} {\bibfnamefont {Kaushik}\ \bibnamefont {Bhattacharya}}, \bibinfo {author} {\bibfnamefont {Andrew}\ \bibnamefont {Stuart}}, \ and\ \bibinfo {author} {\bibfnamefont {Anima}\ \bibnamefont {Anandkumar}},\ }\bibfield  {title} {\enquote {\bibinfo {title} {Fourier neural operator for parametric partial differential equations},}\ }\href@noop {} {\bibfield  {journal} {\bibinfo  {journal} {arXiv preprint arXiv:2010.08895}\ } (\bibinfo {year} {2020}{\natexlab{a}})}\BibitemShut {NoStop}%
\bibitem [{\citenamefont {Kovachki}\ \emph {et~al.}(2023)\citenamefont {Kovachki}, \citenamefont {Li}, \citenamefont {Liu}, \citenamefont {Azizzadenesheli}, \citenamefont {Bhattacharya}, \citenamefont {Stuart},\ and\ \citenamefont {Anandkumar}}]{kovachki2023neural}%
  \BibitemOpen
  \bibfield  {author} {\bibinfo {author} {\bibfnamefont {Nikola}\ \bibnamefont {Kovachki}}, \bibinfo {author} {\bibfnamefont {Zongyi}\ \bibnamefont {Li}}, \bibinfo {author} {\bibfnamefont {Burigede}\ \bibnamefont {Liu}}, \bibinfo {author} {\bibfnamefont {Kamyar}\ \bibnamefont {Azizzadenesheli}}, \bibinfo {author} {\bibfnamefont {Kaushik}\ \bibnamefont {Bhattacharya}}, \bibinfo {author} {\bibfnamefont {Andrew}\ \bibnamefont {Stuart}}, \ and\ \bibinfo {author} {\bibfnamefont {Anima}\ \bibnamefont {Anandkumar}},\ }\bibfield  {title} {\enquote {\bibinfo {title} {Neural operator: Learning maps between function spaces with applications to pdes},}\ }\href@noop {} {\bibfield  {journal} {\bibinfo  {journal} {Journal of Machine Learning Research}\ }\textbf {\bibinfo {volume} {24}},\ \bibinfo {pages} {1--97} (\bibinfo {year} {2023})}\BibitemShut {NoStop}%
\bibitem [{\citenamefont {Li}\ \emph {et~al.}(2020{\natexlab{b}})\citenamefont {Li}, \citenamefont {Kovachki}, \citenamefont {Azizzadenesheli}, \citenamefont {Liu}, \citenamefont {Bhattacharya}, \citenamefont {Stuart},\ and\ \citenamefont {Anandkumar}}]{li2020neural}%
  \BibitemOpen
  \bibfield  {author} {\bibinfo {author} {\bibfnamefont {Zongyi}\ \bibnamefont {Li}}, \bibinfo {author} {\bibfnamefont {Nikola}\ \bibnamefont {Kovachki}}, \bibinfo {author} {\bibfnamefont {Kamyar}\ \bibnamefont {Azizzadenesheli}}, \bibinfo {author} {\bibfnamefont {Burigede}\ \bibnamefont {Liu}}, \bibinfo {author} {\bibfnamefont {Kaushik}\ \bibnamefont {Bhattacharya}}, \bibinfo {author} {\bibfnamefont {Andrew}\ \bibnamefont {Stuart}}, \ and\ \bibinfo {author} {\bibfnamefont {Anima}\ \bibnamefont {Anandkumar}},\ }\bibfield  {title} {\enquote {\bibinfo {title} {Neural operator: Graph kernel network for partial differential equations},}\ }\href@noop {} {\bibfield  {journal} {\bibinfo  {journal} {arXiv preprint arXiv:2003.03485}\ } (\bibinfo {year} {2020}{\natexlab{b}})}\BibitemShut {NoStop}%
\bibitem [{\citenamefont {Li}\ \emph {et~al.}(2022)\citenamefont {Li}, \citenamefont {Meidani},\ and\ \citenamefont {Farimani}}]{li2022transformer}%
  \BibitemOpen
  \bibfield  {author} {\bibinfo {author} {\bibfnamefont {Zijie}\ \bibnamefont {Li}}, \bibinfo {author} {\bibfnamefont {Kazem}\ \bibnamefont {Meidani}}, \ and\ \bibinfo {author} {\bibfnamefont {Amir~Barati}\ \bibnamefont {Farimani}},\ }\bibfield  {title} {\enquote {\bibinfo {title} {Transformer for partial differential equations' operator learning},}\ }\href@noop {} {\bibfield  {journal} {\bibinfo  {journal} {arXiv preprint arXiv:2205.13671}\ } (\bibinfo {year} {2022})}\BibitemShut {NoStop}%
\bibitem [{\citenamefont {Zhu}\ \emph {et~al.}(2024)\citenamefont {Zhu}, \citenamefont {Yin}, \citenamefont {Reeves}, \citenamefont {Yang},\ and\ \citenamefont {Vlcek}}]{zhu2024predicting}%
  \BibitemOpen
  \bibfield  {author} {\bibinfo {author} {\bibfnamefont {Yuanran}\ \bibnamefont {Zhu}}, \bibinfo {author} {\bibfnamefont {Jia}\ \bibnamefont {Yin}}, \bibinfo {author} {\bibfnamefont {Cian~C}\ \bibnamefont {Reeves}}, \bibinfo {author} {\bibfnamefont {Chao}\ \bibnamefont {Yang}}, \ and\ \bibinfo {author} {\bibfnamefont {Vojtech}\ \bibnamefont {Vlcek}},\ }\bibfield  {title} {\enquote {\bibinfo {title} {Predicting nonequilibrium green's function dynamics and photoemission spectra via nonlinear integral operator learning},}\ }\href@noop {} {\bibfield  {journal} {\bibinfo  {journal} {arXiv preprint arXiv:2407.09773}\ } (\bibinfo {year} {2024})}\BibitemShut {NoStop}%
\bibitem [{\citenamefont {Reeves}\ \emph {et~al.}(2024)\citenamefont {Reeves}, \citenamefont {Harsha}, \citenamefont {Shee}, \citenamefont {Zhu}, \citenamefont {Yang}, \citenamefont {Whaley}, \citenamefont {Zgid},\ and\ \citenamefont {Vlcek}}]{reeves2024performance}%
  \BibitemOpen
  \bibfield  {author} {\bibinfo {author} {\bibfnamefont {Cian~C}\ \bibnamefont {Reeves}}, \bibinfo {author} {\bibfnamefont {Gaurav}\ \bibnamefont {Harsha}}, \bibinfo {author} {\bibfnamefont {Avijit}\ \bibnamefont {Shee}}, \bibinfo {author} {\bibfnamefont {Yuanran}\ \bibnamefont {Zhu}}, \bibinfo {author} {\bibfnamefont {Chao}\ \bibnamefont {Yang}}, \bibinfo {author} {\bibfnamefont {K~Birgitta}\ \bibnamefont {Whaley}}, \bibinfo {author} {\bibfnamefont {Dominika}\ \bibnamefont {Zgid}}, \ and\ \bibinfo {author} {\bibfnamefont {Vojtech}\ \bibnamefont {Vlcek}},\ }\bibfield  {title} {\enquote {\bibinfo {title} {Performance of wave function and green's functions based methods for non equilibrium many-body dynamics},}\ }\href@noop {} {\bibfield  {journal} {\bibinfo  {journal} {arXiv preprint arXiv:2405.08814}\ } (\bibinfo {year} {2024})}\BibitemShut {NoStop}%
\bibitem [{\citenamefont {Brunton}\ and\ \citenamefont {Kutz}(2022)}]{brunton2022data}%
  \BibitemOpen
  \bibfield  {author} {\bibinfo {author} {\bibfnamefont {Steven~L}\ \bibnamefont {Brunton}}\ and\ \bibinfo {author} {\bibfnamefont {J~Nathan}\ \bibnamefont {Kutz}},\ }\href@noop {} {\emph {\bibinfo {title} {Data-driven science and engineering: Machine learning, dynamical systems, and control}}}\ (\bibinfo  {publisher} {Cambridge University Press},\ \bibinfo {year} {2022})\BibitemShut {NoStop}%
\bibitem [{\citenamefont {Shwartz-Ziv}\ and\ \citenamefont {Tishby}(2017)}]{shwartz2017opening}%
  \BibitemOpen
  \bibfield  {author} {\bibinfo {author} {\bibfnamefont {Ravid}\ \bibnamefont {Shwartz-Ziv}}\ and\ \bibinfo {author} {\bibfnamefont {Naftali}\ \bibnamefont {Tishby}},\ }\bibfield  {title} {\enquote {\bibinfo {title} {Opening the black box of deep neural networks via information},}\ }\href@noop {} {\bibfield  {journal} {\bibinfo  {journal} {arXiv preprint arXiv:1703.00810}\ } (\bibinfo {year} {2017})}\BibitemShut {NoStop}%
\bibitem [{\citenamefont {Kremp}\ \emph {et~al.}(2005)\citenamefont {Kremp}, \citenamefont {Schlanges},\ and\ \citenamefont {Kraeft}}]{kremp2005quantum}%
  \BibitemOpen
  \bibfield  {author} {\bibinfo {author} {\bibfnamefont {Dietrich}\ \bibnamefont {Kremp}}, \bibinfo {author} {\bibfnamefont {Manfred}\ \bibnamefont {Schlanges}}, \ and\ \bibinfo {author} {\bibfnamefont {Wolf-Dietrich}\ \bibnamefont {Kraeft}},\ }\href@noop {} {\emph {\bibinfo {title} {Quantum statistics of nonideal plasmas}}},\ Vol.~\bibinfo {volume} {25}\ (\bibinfo  {publisher} {Springer Science \& Business Media},\ \bibinfo {year} {2005})\BibitemShut {NoStop}%
\bibitem [{\citenamefont {Kadanoff}(2018)}]{kadanoff2018quantum}%
  \BibitemOpen
  \bibfield  {author} {\bibinfo {author} {\bibfnamefont {Leo~P}\ \bibnamefont {Kadanoff}},\ }\href@noop {} {\emph {\bibinfo {title} {Quantum statistical mechanics}}}\ (\bibinfo  {publisher} {CRC Press},\ \bibinfo {year} {2018})\BibitemShut {NoStop}%
\bibitem [{\citenamefont {Kadanoff}\ \emph {et~al.}(1962)\citenamefont {Kadanoff}, \citenamefont {Baym},\ and\ \citenamefont {Mechanics}}]{kadanoff1962green}%
  \BibitemOpen
  \bibfield  {author} {\bibinfo {author} {\bibfnamefont {LP}~\bibnamefont {Kadanoff}}, \bibinfo {author} {\bibfnamefont {G}~\bibnamefont {Baym}}, \ and\ \bibinfo {author} {\bibfnamefont {Quantum~Statistical}\ \bibnamefont {Mechanics}},\ }\bibfield  {title} {\enquote {\bibinfo {title} {Green’s function methods in equilibrium and nonequilibrium problems},}\ }\href@noop {} {\bibfield  {journal} {\bibinfo  {journal} {Frontiers in Physics (Benjamin, New York, 1962)}\ } (\bibinfo {year} {1962})}\BibitemShut {NoStop}%
\bibitem [{\citenamefont {Hedin}(1965)}]{hedin1965new}%
  \BibitemOpen
  \bibfield  {author} {\bibinfo {author} {\bibfnamefont {Lars}\ \bibnamefont {Hedin}},\ }\bibfield  {title} {\enquote {\bibinfo {title} {New method for calculating the one-particle green's function with application to the electron-gas problem},}\ }\href@noop {} {\bibfield  {journal} {\bibinfo  {journal} {Physical Review}\ }\textbf {\bibinfo {volume} {139}},\ \bibinfo {pages} {A796} (\bibinfo {year} {1965})}\BibitemShut {NoStop}%
\bibitem [{\citenamefont {Hedin}\ and\ \citenamefont {Lundqvist}(1970)}]{hedin1970effects}%
  \BibitemOpen
  \bibfield  {author} {\bibinfo {author} {\bibfnamefont {Lars}\ \bibnamefont {Hedin}}\ and\ \bibinfo {author} {\bibfnamefont {Stig}\ \bibnamefont {Lundqvist}},\ }\bibfield  {title} {\enquote {\bibinfo {title} {Effects of electron-electron and electron-phonon interactions on the one-electron states of solids},}\ }in\ \href@noop {} {\emph {\bibinfo {booktitle} {Solid state physics}}},\ Vol.~\bibinfo {volume} {23}\ (\bibinfo  {publisher} {Elsevier},\ \bibinfo {year} {1970})\ pp.\ \bibinfo {pages} {1--181}\BibitemShut {NoStop}%
\bibitem [{\citenamefont {Hybertsen}\ and\ \citenamefont {Louie}(1986{\natexlab{a}})}]{hybertsen1986electron}%
  \BibitemOpen
  \bibfield  {author} {\bibinfo {author} {\bibfnamefont {Mark~S}\ \bibnamefont {Hybertsen}}\ and\ \bibinfo {author} {\bibfnamefont {Steven~G}\ \bibnamefont {Louie}},\ }\bibfield  {title} {\enquote {\bibinfo {title} {Electron correlation in semiconductors and insulators: Band gaps and quasiparticle energies},}\ }\href@noop {} {\bibfield  {journal} {\bibinfo  {journal} {Physical Review B}\ }\textbf {\bibinfo {volume} {34}},\ \bibinfo {pages} {5390} (\bibinfo {year} {1986}{\natexlab{a}})}\BibitemShut {NoStop}%
\bibitem [{\citenamefont {Adler}(1962)}]{adler1962quantum}%
  \BibitemOpen
  \bibfield  {author} {\bibinfo {author} {\bibfnamefont {Stephen~L}\ \bibnamefont {Adler}},\ }\bibfield  {title} {\enquote {\bibinfo {title} {Quantum theory of the dielectric constant in real solids},}\ }\href@noop {} {\bibfield  {journal} {\bibinfo  {journal} {Physical Review}\ }\textbf {\bibinfo {volume} {126}},\ \bibinfo {pages} {413} (\bibinfo {year} {1962})}\BibitemShut {NoStop}%
\bibitem [{\citenamefont {Wiser}(1963)}]{wiser1963dielectric}%
  \BibitemOpen
  \bibfield  {author} {\bibinfo {author} {\bibfnamefont {Nathan}\ \bibnamefont {Wiser}},\ }\bibfield  {title} {\enquote {\bibinfo {title} {Dielectric constant with local field effects included},}\ }\href@noop {} {\bibfield  {journal} {\bibinfo  {journal} {Physical Review}\ }\textbf {\bibinfo {volume} {129}},\ \bibinfo {pages} {62} (\bibinfo {year} {1963})}\BibitemShut {NoStop}%
\bibitem [{\citenamefont {Ceperley}\ and\ \citenamefont {Alder}(1980)}]{ceperley1980ground}%
  \BibitemOpen
  \bibfield  {author} {\bibinfo {author} {\bibfnamefont {David~M}\ \bibnamefont {Ceperley}}\ and\ \bibinfo {author} {\bibfnamefont {Berni~J}\ \bibnamefont {Alder}},\ }\bibfield  {title} {\enquote {\bibinfo {title} {Ground state of the electron gas by a stochastic method},}\ }\href@noop {} {\bibfield  {journal} {\bibinfo  {journal} {Physical review letters}\ }\textbf {\bibinfo {volume} {45}},\ \bibinfo {pages} {566} (\bibinfo {year} {1980})}\BibitemShut {NoStop}%
\bibitem [{\citenamefont {Aversa}\ and\ \citenamefont {Sipe}(1995)}]{aversa1995nonlinear}%
  \BibitemOpen
  \bibfield  {author} {\bibinfo {author} {\bibfnamefont {Claudio}\ \bibnamefont {Aversa}}\ and\ \bibinfo {author} {\bibfnamefont {John~E}\ \bibnamefont {Sipe}},\ }\bibfield  {title} {\enquote {\bibinfo {title} {Nonlinear optical susceptibilities of semiconductors: Results with a length-gauge analysis},}\ }\href@noop {} {\bibfield  {journal} {\bibinfo  {journal} {Physical Review B}\ }\textbf {\bibinfo {volume} {52}},\ \bibinfo {pages} {14636} (\bibinfo {year} {1995})}\BibitemShut {NoStop}%
\bibitem [{\citenamefont {Virk}\ and\ \citenamefont {Sipe}(2007{\natexlab{a}})}]{virk2007semiconductor}%
  \BibitemOpen
  \bibfield  {author} {\bibinfo {author} {\bibfnamefont {Kuljit~S}\ \bibnamefont {Virk}}\ and\ \bibinfo {author} {\bibfnamefont {JE}~\bibnamefont {Sipe}},\ }\bibfield  {title} {\enquote {\bibinfo {title} {Semiconductor optics in length gauge: A general numerical approach},}\ }\href@noop {} {\bibfield  {journal} {\bibinfo  {journal} {Physical Review B—Condensed Matter and Materials Physics}\ }\textbf {\bibinfo {volume} {76}},\ \bibinfo {pages} {035213} (\bibinfo {year} {2007}{\natexlab{a}})}\BibitemShut {NoStop}%
\bibitem [{\citenamefont {Rocca}\ \emph {et~al.}(2012)\citenamefont {Rocca}, \citenamefont {Ping}, \citenamefont {Gebauer},\ and\ \citenamefont {Galli}}]{rocca2012solution}%
  \BibitemOpen
  \bibfield  {author} {\bibinfo {author} {\bibfnamefont {Dario}\ \bibnamefont {Rocca}}, \bibinfo {author} {\bibfnamefont {Yuan}\ \bibnamefont {Ping}}, \bibinfo {author} {\bibfnamefont {Ralph}\ \bibnamefont {Gebauer}}, \ and\ \bibinfo {author} {\bibfnamefont {Giulia}\ \bibnamefont {Galli}},\ }\bibfield  {title} {\enquote {\bibinfo {title} {Solution of the bethe-salpeter equation without empty electronic states: Application to the absorption spectra of bulk systems},}\ }\href@noop {} {\bibfield  {journal} {\bibinfo  {journal} {Physical Review B—Condensed Matter and Materials Physics}\ }\textbf {\bibinfo {volume} {85}},\ \bibinfo {pages} {045116} (\bibinfo {year} {2012})}\BibitemShut {NoStop}%
\bibitem [{\citenamefont {Rohlfing}\ and\ \citenamefont {Louie}(2000{\natexlab{a}})}]{rohlfing2000electron}%
  \BibitemOpen
  \bibfield  {author} {\bibinfo {author} {\bibfnamefont {Michael}\ \bibnamefont {Rohlfing}}\ and\ \bibinfo {author} {\bibfnamefont {Steven~G}\ \bibnamefont {Louie}},\ }\bibfield  {title} {\enquote {\bibinfo {title} {Electron-hole excitations and optical spectra from first principles},}\ }\href@noop {} {\bibfield  {journal} {\bibinfo  {journal} {Physical Review B}\ }\textbf {\bibinfo {volume} {62}},\ \bibinfo {pages} {4927} (\bibinfo {year} {2000}{\natexlab{a}})}\BibitemShut {NoStop}%
\bibitem [{\citenamefont {Albrecht}\ \emph {et~al.}(1998)\citenamefont {Albrecht}, \citenamefont {Reining}, \citenamefont {Del~Sole},\ and\ \citenamefont {Onida}}]{albrecht1998ab}%
  \BibitemOpen
  \bibfield  {author} {\bibinfo {author} {\bibfnamefont {Stefan}\ \bibnamefont {Albrecht}}, \bibinfo {author} {\bibfnamefont {Lucia}\ \bibnamefont {Reining}}, \bibinfo {author} {\bibfnamefont {Rodolfo}\ \bibnamefont {Del~Sole}}, \ and\ \bibinfo {author} {\bibfnamefont {Giovanni}\ \bibnamefont {Onida}},\ }\bibfield  {title} {\enquote {\bibinfo {title} {Ab initio calculation of excitonic effects in the optical spectra of semiconductors},}\ }\href@noop {} {\bibfield  {journal} {\bibinfo  {journal} {Physical review letters}\ }\textbf {\bibinfo {volume} {80}},\ \bibinfo {pages} {4510} (\bibinfo {year} {1998})}\BibitemShut {NoStop}%
\bibitem [{\citenamefont {Deslippe}\ \emph {et~al.}(2012{\natexlab{a}})\citenamefont {Deslippe}, \citenamefont {Samsonidze}, \citenamefont {Strubbe}, \citenamefont {Jain}, \citenamefont {Cohen},\ and\ \citenamefont {Louie}}]{BGW1}%
  \BibitemOpen
  \bibfield  {author} {\bibinfo {author} {\bibfnamefont {Jack}\ \bibnamefont {Deslippe}}, \bibinfo {author} {\bibfnamefont {Georgy}\ \bibnamefont {Samsonidze}}, \bibinfo {author} {\bibfnamefont {David~A.}\ \bibnamefont {Strubbe}}, \bibinfo {author} {\bibfnamefont {Manish}\ \bibnamefont {Jain}}, \bibinfo {author} {\bibfnamefont {Marvin~L.}\ \bibnamefont {Cohen}}, \ and\ \bibinfo {author} {\bibfnamefont {Steven~G.}\ \bibnamefont {Louie}},\ }\bibfield  {title} {\enquote {\bibinfo {title} {Berkeleygw: A massively parallel computer package for the calculation of the quasiparticle and optical properties of materials and nanostructures},}\ }\href {\doibase https://doi.org/10.1016/j.cpc.2011.12.006} {\bibfield  {journal} {\bibinfo  {journal} {Computer Physics Communications}\ }\textbf {\bibinfo {volume} {183}},\ \bibinfo {pages} {1269--1289} (\bibinfo {year} {2012}{\natexlab{a}})}\BibitemShut {NoStop}%
\bibitem [{\citenamefont {Rohlfing}\ and\ \citenamefont {Louie}(2000{\natexlab{b}})}]{Rohlfing2000}%
  \BibitemOpen
  \bibfield  {author} {\bibinfo {author} {\bibfnamefont {Michael}\ \bibnamefont {Rohlfing}}\ and\ \bibinfo {author} {\bibfnamefont {Steven~G.}\ \bibnamefont {Louie}},\ }\bibfield  {title} {\enquote {\bibinfo {title} {Electron-hole excitations and optical spectra from first principles},}\ }\href {\doibase 10.1103/PhysRevB.62.4927} {\bibfield  {journal} {\bibinfo  {journal} {Phys. Rev. B}\ }\textbf {\bibinfo {volume} {62}},\ \bibinfo {pages} {4927--4944} (\bibinfo {year} {2000}{\natexlab{b}})}\BibitemShut {NoStop}%
\bibitem [{\citenamefont {Strinati}(1988)}]{strinati1988application}%
  \BibitemOpen
  \bibfield  {author} {\bibinfo {author} {\bibfnamefont {Giancarlo}\ \bibnamefont {Strinati}},\ }\bibfield  {title} {\enquote {\bibinfo {title} {Application of the green’s functions method to the study of the optical properties of semiconductors},}\ }\href@noop {} {\bibfield  {journal} {\bibinfo  {journal} {La Rivista del Nuovo Cimento (1978-1999)}\ }\textbf {\bibinfo {volume} {11}},\ \bibinfo {pages} {1--86} (\bibinfo {year} {1988})}\BibitemShut {NoStop}%
\bibitem [{\citenamefont {Giannozzi}\ \emph {et~al.}(2009)\citenamefont {Giannozzi}, \citenamefont {Baroni}, \citenamefont {Bonini}, \citenamefont {Calandra}, \citenamefont {Car}, \citenamefont {Cavazzoni}, \citenamefont {Ceresoli}, \citenamefont {Chiarotti}, \citenamefont {Cococcioni}, \citenamefont {Dabo}, \citenamefont {Corso}, \citenamefont {de~Gironcoli}, \citenamefont {Fabris}, \citenamefont {Fratesi}, \citenamefont {Gebauer}, \citenamefont {Gerstmann}, \citenamefont {Gougoussis}, \citenamefont {Kokalj}, \citenamefont {Lazzeri}, \citenamefont {Martin-Samos}, \citenamefont {Marzari}, \citenamefont {Mauri}, \citenamefont {Mazzarello}, \citenamefont {Paolini}, \citenamefont {Pasquarello}, \citenamefont {Paulatto}, \citenamefont {Sbraccia}, \citenamefont {Scandolo}, \citenamefont {Sclauzero}, \citenamefont {Seitsonen}, \citenamefont {Smogunov}, \citenamefont {Umari},\ and\ \citenamefont {Wentzcovitch}}]{Giannozzi_2009}%
  \BibitemOpen
  \bibfield  {author} {\bibinfo {author} {\bibfnamefont {Paolo}\ \bibnamefont {Giannozzi}}, \bibinfo {author} {\bibfnamefont {Stefano}\ \bibnamefont {Baroni}}, \bibinfo {author} {\bibfnamefont {Nicola}\ \bibnamefont {Bonini}}, \bibinfo {author} {\bibfnamefont {Matteo}\ \bibnamefont {Calandra}}, \bibinfo {author} {\bibfnamefont {Roberto}\ \bibnamefont {Car}}, \bibinfo {author} {\bibfnamefont {Carlo}\ \bibnamefont {Cavazzoni}}, \bibinfo {author} {\bibfnamefont {Davide}\ \bibnamefont {Ceresoli}}, \bibinfo {author} {\bibfnamefont {Guido~L}\ \bibnamefont {Chiarotti}}, \bibinfo {author} {\bibfnamefont {Matteo}\ \bibnamefont {Cococcioni}}, \bibinfo {author} {\bibfnamefont {Ismaila}\ \bibnamefont {Dabo}}, \bibinfo {author} {\bibfnamefont {Andrea~Dal}\ \bibnamefont {Corso}}, \bibinfo {author} {\bibfnamefont {Stefano}\ \bibnamefont {de~Gironcoli}}, \bibinfo {author} {\bibfnamefont {Stefano}\ \bibnamefont {Fabris}}, \bibinfo {author} {\bibfnamefont {Guido}\ \bibnamefont {Fratesi}}, \bibinfo {author} {\bibfnamefont
  {Ralph}\ \bibnamefont {Gebauer}}, \bibinfo {author} {\bibfnamefont {Uwe}\ \bibnamefont {Gerstmann}}, \bibinfo {author} {\bibfnamefont {Christos}\ \bibnamefont {Gougoussis}}, \bibinfo {author} {\bibfnamefont {Anton}\ \bibnamefont {Kokalj}}, \bibinfo {author} {\bibfnamefont {Michele}\ \bibnamefont {Lazzeri}}, \bibinfo {author} {\bibfnamefont {Layla}\ \bibnamefont {Martin-Samos}}, \bibinfo {author} {\bibfnamefont {Nicola}\ \bibnamefont {Marzari}}, \bibinfo {author} {\bibfnamefont {Francesco}\ \bibnamefont {Mauri}}, \bibinfo {author} {\bibfnamefont {Riccardo}\ \bibnamefont {Mazzarello}}, \bibinfo {author} {\bibfnamefont {Stefano}\ \bibnamefont {Paolini}}, \bibinfo {author} {\bibfnamefont {Alfredo}\ \bibnamefont {Pasquarello}}, \bibinfo {author} {\bibfnamefont {Lorenzo}\ \bibnamefont {Paulatto}}, \bibinfo {author} {\bibfnamefont {Carlo}\ \bibnamefont {Sbraccia}}, \bibinfo {author} {\bibfnamefont {Sandro}\ \bibnamefont {Scandolo}}, \bibinfo {author} {\bibfnamefont {Gabriele}\ \bibnamefont {Sclauzero}}, \bibinfo
  {author} {\bibfnamefont {Ari~P}\ \bibnamefont {Seitsonen}}, \bibinfo {author} {\bibfnamefont {Alexander}\ \bibnamefont {Smogunov}}, \bibinfo {author} {\bibfnamefont {Paolo}\ \bibnamefont {Umari}}, \ and\ \bibinfo {author} {\bibfnamefont {Renata~M}\ \bibnamefont {Wentzcovitch}},\ }\bibfield  {title} {\enquote {\bibinfo {title} {Quantum espresso: a modular and open-source software project for quantum simulations of materials},}\ }\href {\doibase 10.1088/0953-8984/21/39/395502} {\bibfield  {journal} {\bibinfo  {journal} {Journal of Physics: Condensed Matter}\ }\textbf {\bibinfo {volume} {21}},\ \bibinfo {pages} {395502} (\bibinfo {year} {2009})}\BibitemShut {NoStop}%
\bibitem [{\citenamefont {Kohn}\ and\ \citenamefont {Sham}(1965)}]{kohn1965self}%
  \BibitemOpen
  \bibfield  {author} {\bibinfo {author} {\bibfnamefont {Walter}\ \bibnamefont {Kohn}}\ and\ \bibinfo {author} {\bibfnamefont {Lu~Jeu}\ \bibnamefont {Sham}},\ }\bibfield  {title} {\enquote {\bibinfo {title} {Self-consistent equations including exchange and correlation effects},}\ }\href@noop {} {\bibfield  {journal} {\bibinfo  {journal} {Physical review}\ }\textbf {\bibinfo {volume} {140}},\ \bibinfo {pages} {A1133} (\bibinfo {year} {1965})}\BibitemShut {NoStop}%
\bibitem [{\citenamefont {Alvertis}\ \emph {et~al.}(2023)\citenamefont {Alvertis}, \citenamefont {Champagne}, \citenamefont {Del~Ben}, \citenamefont {da~Jornada}, \citenamefont {Qiu}, \citenamefont {Filip},\ and\ \citenamefont {Neaton}}]{alvertis2023importance}%
  \BibitemOpen
  \bibfield  {author} {\bibinfo {author} {\bibfnamefont {Antonios~M}\ \bibnamefont {Alvertis}}, \bibinfo {author} {\bibfnamefont {Aur{\'e}lie}\ \bibnamefont {Champagne}}, \bibinfo {author} {\bibfnamefont {Mauro}\ \bibnamefont {Del~Ben}}, \bibinfo {author} {\bibfnamefont {Felipe~H}\ \bibnamefont {da~Jornada}}, \bibinfo {author} {\bibfnamefont {Diana~Y}\ \bibnamefont {Qiu}}, \bibinfo {author} {\bibfnamefont {Marina~R}\ \bibnamefont {Filip}}, \ and\ \bibinfo {author} {\bibfnamefont {Jeffrey~B}\ \bibnamefont {Neaton}},\ }\bibfield  {title} {\enquote {\bibinfo {title} {Importance of nonuniform brillouin zone sampling for ab initio bethe-salpeter equation calculations of exciton binding energies in crystalline solids},}\ }\href@noop {} {\bibfield  {journal} {\bibinfo  {journal} {Physical Review B}\ }\textbf {\bibinfo {volume} {108}},\ \bibinfo {pages} {235117} (\bibinfo {year} {2023})}\BibitemShut {NoStop}%
\bibitem [{\citenamefont {Hou}\ \emph {et~al.}(2023)\citenamefont {Hou}, \citenamefont {Wang}, \citenamefont {Barker},\ and\ \citenamefont {Qiu}}]{Hou2023Bi2Se3}%
  \BibitemOpen
  \bibfield  {author} {\bibinfo {author} {\bibfnamefont {Bowen}\ \bibnamefont {Hou}}, \bibinfo {author} {\bibfnamefont {Dan}\ \bibnamefont {Wang}}, \bibinfo {author} {\bibfnamefont {Bradford~A.}\ \bibnamefont {Barker}}, \ and\ \bibinfo {author} {\bibfnamefont {Diana~Y.}\ \bibnamefont {Qiu}},\ }\bibfield  {title} {\enquote {\bibinfo {title} {Exchange-driven intermixing of bulk and topological surface states by chiral excitons in ${\mathrm{bi}}_{2}{\mathrm{se}}_{3}$},}\ }\href {\doibase 10.1103/PhysRevLett.130.216402} {\bibfield  {journal} {\bibinfo  {journal} {Phys. Rev. Lett.}\ }\textbf {\bibinfo {volume} {130}},\ \bibinfo {pages} {216402} (\bibinfo {year} {2023})}\BibitemShut {NoStop}%
\bibitem [{\citenamefont {Li}\ \emph {et~al.}(2017)\citenamefont {Li}, \citenamefont {Kim}, \citenamefont {Jin}, \citenamefont {Ye}, \citenamefont {Qiu}, \citenamefont {Da~Jornada}, \citenamefont {Shi}, \citenamefont {Chen}, \citenamefont {Zhang}, \citenamefont {Yang} \emph {et~al.}}]{li2017direct}%
  \BibitemOpen
  \bibfield  {author} {\bibinfo {author} {\bibfnamefont {Likai}\ \bibnamefont {Li}}, \bibinfo {author} {\bibfnamefont {Jonghwan}\ \bibnamefont {Kim}}, \bibinfo {author} {\bibfnamefont {Chenhao}\ \bibnamefont {Jin}}, \bibinfo {author} {\bibfnamefont {Guo~Jun}\ \bibnamefont {Ye}}, \bibinfo {author} {\bibfnamefont {Diana~Y}\ \bibnamefont {Qiu}}, \bibinfo {author} {\bibfnamefont {Felipe~H}\ \bibnamefont {Da~Jornada}}, \bibinfo {author} {\bibfnamefont {Zhiwen}\ \bibnamefont {Shi}}, \bibinfo {author} {\bibfnamefont {Long}\ \bibnamefont {Chen}}, \bibinfo {author} {\bibfnamefont {Zuocheng}\ \bibnamefont {Zhang}}, \bibinfo {author} {\bibfnamefont {Fangyuan}\ \bibnamefont {Yang}},  \emph {et~al.},\ }\bibfield  {title} {\enquote {\bibinfo {title} {Direct observation of the layer-dependent electronic structure in phosphorene},}\ }\href@noop {} {\bibfield  {journal} {\bibinfo  {journal} {Nature nanotechnology}\ }\textbf {\bibinfo {volume} {12}},\ \bibinfo {pages} {21--25} (\bibinfo {year} {2017})}\BibitemShut {NoStop}%
\bibitem [{\citenamefont {Wu}\ \emph {et~al.}(2024)\citenamefont {Wu}, \citenamefont {Hou}, \citenamefont {Li}, \citenamefont {He},\ and\ \citenamefont {Qiu}}]{Wu2024WTe2}%
  \BibitemOpen
  \bibfield  {author} {\bibinfo {author} {\bibfnamefont {Jinyuan}\ \bibnamefont {Wu}}, \bibinfo {author} {\bibfnamefont {Bowen}\ \bibnamefont {Hou}}, \bibinfo {author} {\bibfnamefont {Wenxin}\ \bibnamefont {Li}}, \bibinfo {author} {\bibfnamefont {Yu}~\bibnamefont {He}}, \ and\ \bibinfo {author} {\bibfnamefont {Diana~Y.}\ \bibnamefont {Qiu}},\ }\bibfield  {title} {\enquote {\bibinfo {title} {Quasiparticle and excitonic properties of monolayer $1{T}^{\ensuremath{'}}$ ${\mathrm{wte}}_{2}$ within many-body perturbation theory},}\ }\href {\doibase 10.1103/PhysRevB.110.075133} {\bibfield  {journal} {\bibinfo  {journal} {Phys. Rev. B}\ }\textbf {\bibinfo {volume} {110}},\ \bibinfo {pages} {075133} (\bibinfo {year} {2024})}\BibitemShut {NoStop}%
\bibitem [{\citenamefont {Qiu}\ \emph {et~al.}(2016)\citenamefont {Qiu}, \citenamefont {da~Jornada},\ and\ \citenamefont {Louie}}]{Qiu2012MoS2}%
  \BibitemOpen
  \bibfield  {author} {\bibinfo {author} {\bibfnamefont {Diana~Y.}\ \bibnamefont {Qiu}}, \bibinfo {author} {\bibfnamefont {Felipe~H.}\ \bibnamefont {da~Jornada}}, \ and\ \bibinfo {author} {\bibfnamefont {Steven~G.}\ \bibnamefont {Louie}},\ }\bibfield  {title} {\enquote {\bibinfo {title} {Screening and many-body effects in two-dimensional crystals: Monolayer ${\mathrm{mos}}_{2}$},}\ }\href {\doibase 10.1103/PhysRevB.93.235435} {\bibfield  {journal} {\bibinfo  {journal} {Phys. Rev. B}\ }\textbf {\bibinfo {volume} {93}},\ \bibinfo {pages} {235435} (\bibinfo {year} {2016})}\BibitemShut {NoStop}%
\bibitem [{\citenamefont {Qiu}\ \emph {et~al.}(2013)\citenamefont {Qiu}, \citenamefont {Da~Jornada},\ and\ \citenamefont {Louie}}]{qiu2013optical}%
  \BibitemOpen
  \bibfield  {author} {\bibinfo {author} {\bibfnamefont {Diana~Y}\ \bibnamefont {Qiu}}, \bibinfo {author} {\bibfnamefont {Felipe~H}\ \bibnamefont {Da~Jornada}}, \ and\ \bibinfo {author} {\bibfnamefont {Steven~G}\ \bibnamefont {Louie}},\ }\bibfield  {title} {\enquote {\bibinfo {title} {Optical spectrum of mos 2: many-body effects and diversity of exciton states},}\ }\href@noop {} {\bibfield  {journal} {\bibinfo  {journal} {Physical review letters}\ }\textbf {\bibinfo {volume} {111}},\ \bibinfo {pages} {216805} (\bibinfo {year} {2013})}\BibitemShut {NoStop}%
\bibitem [{\citenamefont {Cao}\ \emph {et~al.}(2016)\citenamefont {Cao}, \citenamefont {Li}, \citenamefont {Qiu},\ and\ \citenamefont {Louie}}]{cao2016gate}%
  \BibitemOpen
  \bibfield  {author} {\bibinfo {author} {\bibfnamefont {Ting}\ \bibnamefont {Cao}}, \bibinfo {author} {\bibfnamefont {Zhenglu}\ \bibnamefont {Li}}, \bibinfo {author} {\bibfnamefont {Diana~Y}\ \bibnamefont {Qiu}}, \ and\ \bibinfo {author} {\bibfnamefont {Steven~G}\ \bibnamefont {Louie}},\ }\bibfield  {title} {\enquote {\bibinfo {title} {Gate switchable transport and optical anisotropy in 90 twisted bilayer black phosphorus},}\ }\href@noop {} {\bibfield  {journal} {\bibinfo  {journal} {Nano letters}\ }\textbf {\bibinfo {volume} {16}},\ \bibinfo {pages} {5542--5546} (\bibinfo {year} {2016})}\BibitemShut {NoStop}%
\bibitem [{\citenamefont {Qiu}\ \emph {et~al.}(2017)\citenamefont {Qiu}, \citenamefont {da~Jornada},\ and\ \citenamefont {Louie}}]{qiu2017environmental}%
  \BibitemOpen
  \bibfield  {author} {\bibinfo {author} {\bibfnamefont {Diana~Y}\ \bibnamefont {Qiu}}, \bibinfo {author} {\bibfnamefont {Felipe~H}\ \bibnamefont {da~Jornada}}, \ and\ \bibinfo {author} {\bibfnamefont {Steven~G}\ \bibnamefont {Louie}},\ }\bibfield  {title} {\enquote {\bibinfo {title} {Environmental screening effects in 2d materials: renormalization of the bandgap, electronic structure, and optical spectra of few-layer black phosphorus},}\ }\href@noop {} {\bibfield  {journal} {\bibinfo  {journal} {Nano letters}\ }\textbf {\bibinfo {volume} {17}},\ \bibinfo {pages} {4706--4712} (\bibinfo {year} {2017})}\BibitemShut {NoStop}%
\bibitem [{\citenamefont {Qiu}\ \emph {et~al.}(2021{\natexlab{a}})\citenamefont {Qiu}, \citenamefont {da~Jornada},\ and\ \citenamefont {Louie}}]{Qiu2012Sapprox}%
  \BibitemOpen
  \bibfield  {author} {\bibinfo {author} {\bibfnamefont {Diana~Y.}\ \bibnamefont {Qiu}}, \bibinfo {author} {\bibfnamefont {Felipe~H.}\ \bibnamefont {da~Jornada}}, \ and\ \bibinfo {author} {\bibfnamefont {Steven~G.}\ \bibnamefont {Louie}},\ }\bibfield  {title} {\enquote {\bibinfo {title} {Solving the bethe-salpeter equation on a subspace: Approximations and consequences for low-dimensional materials},}\ }\href {\doibase 10.1103/PhysRevB.103.045117} {\bibfield  {journal} {\bibinfo  {journal} {Phys. Rev. B}\ }\textbf {\bibinfo {volume} {103}},\ \bibinfo {pages} {045117} (\bibinfo {year} {2021}{\natexlab{a}})}\BibitemShut {NoStop}%
\bibitem [{\citenamefont {Perdew}\ \emph {et~al.}(1996)\citenamefont {Perdew}, \citenamefont {Burke},\ and\ \citenamefont {Ernzerhof}}]{PBE1}%
  \BibitemOpen
  \bibfield  {author} {\bibinfo {author} {\bibfnamefont {John~P.}\ \bibnamefont {Perdew}}, \bibinfo {author} {\bibfnamefont {Kieron}\ \bibnamefont {Burke}}, \ and\ \bibinfo {author} {\bibfnamefont {Matthias}\ \bibnamefont {Ernzerhof}},\ }\bibfield  {title} {\enquote {\bibinfo {title} {Generalized gradient approximation made simple},}\ }\href {\doibase 10.1103/PhysRevLett.77.3865} {\bibfield  {journal} {\bibinfo  {journal} {Phys. Rev. Lett.}\ }\textbf {\bibinfo {volume} {77}},\ \bibinfo {pages} {3865--3868} (\bibinfo {year} {1996})}\BibitemShut {NoStop}%
\bibitem [{\citenamefont {Hybertsen}\ and\ \citenamefont {Louie}(1986{\natexlab{b}})}]{BGW2}%
  \BibitemOpen
  \bibfield  {author} {\bibinfo {author} {\bibfnamefont {Mark~S.}\ \bibnamefont {Hybertsen}}\ and\ \bibinfo {author} {\bibfnamefont {Steven~G.}\ \bibnamefont {Louie}},\ }\bibfield  {title} {\enquote {\bibinfo {title} {Electron correlation in semiconductors and insulators: Band gaps and quasiparticle energies},}\ }\href {\doibase 10.1103/PhysRevB.34.5390} {\bibfield  {journal} {\bibinfo  {journal} {Phys. Rev. B}\ }\textbf {\bibinfo {volume} {34}},\ \bibinfo {pages} {5390--5413} (\bibinfo {year} {1986}{\natexlab{b}})}\BibitemShut {NoStop}%
\bibitem [{\citenamefont {Rohlfing}\ and\ \citenamefont {Louie}(2000{\natexlab{c}})}]{BGW3}%
  \BibitemOpen
  \bibfield  {author} {\bibinfo {author} {\bibfnamefont {Michael}\ \bibnamefont {Rohlfing}}\ and\ \bibinfo {author} {\bibfnamefont {Steven~G.}\ \bibnamefont {Louie}},\ }\bibfield  {title} {\enquote {\bibinfo {title} {Electron-hole excitations and optical spectra from first principles},}\ }\href {\doibase 10.1103/PhysRevB.62.4927} {\bibfield  {journal} {\bibinfo  {journal} {Phys. Rev. B}\ }\textbf {\bibinfo {volume} {62}},\ \bibinfo {pages} {4927--4944} (\bibinfo {year} {2000}{\natexlab{c}})}\BibitemShut {NoStop}%
\bibitem [{\citenamefont {Souza}\ \emph {et~al.}(2004)\citenamefont {Souza}, \citenamefont {\'I\~niguez},\ and\ \citenamefont {Vanderbilt}}]{localgauge1}%
  \BibitemOpen
  \bibfield  {author} {\bibinfo {author} {\bibfnamefont {Ivo}\ \bibnamefont {Souza}}, \bibinfo {author} {\bibfnamefont {Jorge}\ \bibnamefont {\'I\~niguez}}, \ and\ \bibinfo {author} {\bibfnamefont {David}\ \bibnamefont {Vanderbilt}},\ }\bibfield  {title} {\enquote {\bibinfo {title} {Dynamics of berry-phase polarization in time-dependent electric fields},}\ }\href {\doibase 10.1103/PhysRevB.69.085106} {\bibfield  {journal} {\bibinfo  {journal} {Phys. Rev. B}\ }\textbf {\bibinfo {volume} {69}},\ \bibinfo {pages} {085106} (\bibinfo {year} {2004})}\BibitemShut {NoStop}%
\bibitem [{\citenamefont {Virk}\ and\ \citenamefont {Sipe}(2007{\natexlab{b}})}]{localgauge2}%
  \BibitemOpen
  \bibfield  {author} {\bibinfo {author} {\bibfnamefont {Kuljit~S.}\ \bibnamefont {Virk}}\ and\ \bibinfo {author} {\bibfnamefont {J.~E.}\ \bibnamefont {Sipe}},\ }\bibfield  {title} {\enquote {\bibinfo {title} {Semiconductor optics in length gauge: A general numerical approach},}\ }\href {\doibase 10.1103/PhysRevB.76.035213} {\bibfield  {journal} {\bibinfo  {journal} {Phys. Rev. B}\ }\textbf {\bibinfo {volume} {76}},\ \bibinfo {pages} {035213} (\bibinfo {year} {2007}{\natexlab{b}})}\BibitemShut {NoStop}%
\bibitem [{\citenamefont {Marzari}\ and\ \citenamefont {Vanderbilt}(1997)}]{localgauge3}%
  \BibitemOpen
  \bibfield  {author} {\bibinfo {author} {\bibfnamefont {Nicola}\ \bibnamefont {Marzari}}\ and\ \bibinfo {author} {\bibfnamefont {David}\ \bibnamefont {Vanderbilt}},\ }\bibfield  {title} {\enquote {\bibinfo {title} {Maximally localized generalized wannier functions for composite energy bands},}\ }\href {\doibase 10.1103/PhysRevB.56.12847} {\bibfield  {journal} {\bibinfo  {journal} {Phys. Rev. B}\ }\textbf {\bibinfo {volume} {56}},\ \bibinfo {pages} {12847--12865} (\bibinfo {year} {1997})}\BibitemShut {NoStop}%
\bibitem [{\citenamefont {Chan}\ \emph {et~al.}(2021{\natexlab{b}})\citenamefont {Chan}, \citenamefont {Qiu}, \citenamefont {da~Jornada},\ and\ \citenamefont {Louie}}]{localgauge4}%
  \BibitemOpen
  \bibfield  {author} {\bibinfo {author} {\bibfnamefont {Yang-Hao}\ \bibnamefont {Chan}}, \bibinfo {author} {\bibfnamefont {Diana~Y}\ \bibnamefont {Qiu}}, \bibinfo {author} {\bibfnamefont {Felipe~H}\ \bibnamefont {da~Jornada}}, \ and\ \bibinfo {author} {\bibfnamefont {Steven~G}\ \bibnamefont {Louie}},\ }\bibfield  {title} {\enquote {\bibinfo {title} {Giant exciton-enhanced shift currents and direct current conduction with subbandgap photo excitations produced by many-electron interactions},}\ }\href@noop {} {\bibfield  {journal} {\bibinfo  {journal} {Proceedings of the National Academy of Sciences}\ }\textbf {\bibinfo {volume} {118}},\ \bibinfo {pages} {e1906938118} (\bibinfo {year} {2021}{\natexlab{b}})}\BibitemShut {NoStop}%
\bibitem [{\citenamefont {Deslippe}\ \emph {et~al.}(2012{\natexlab{b}})\citenamefont {Deslippe}, \citenamefont {Samsonidze}, \citenamefont {Strubbe}, \citenamefont {Jain}, \citenamefont {Cohen},\ and\ \citenamefont {Louie}}]{deslippe2012berkeleygw}%
  \BibitemOpen
  \bibfield  {author} {\bibinfo {author} {\bibfnamefont {Jack}\ \bibnamefont {Deslippe}}, \bibinfo {author} {\bibfnamefont {Georgy}\ \bibnamefont {Samsonidze}}, \bibinfo {author} {\bibfnamefont {David~A}\ \bibnamefont {Strubbe}}, \bibinfo {author} {\bibfnamefont {Manish}\ \bibnamefont {Jain}}, \bibinfo {author} {\bibfnamefont {Marvin~L}\ \bibnamefont {Cohen}}, \ and\ \bibinfo {author} {\bibfnamefont {Steven~G}\ \bibnamefont {Louie}},\ }\bibfield  {title} {\enquote {\bibinfo {title} {Berkeleygw: A massively parallel computer package for the calculation of the quasiparticle and optical properties of materials and nanostructures},}\ }\href@noop {} {\bibfield  {journal} {\bibinfo  {journal} {Computer Physics Communications}\ }\textbf {\bibinfo {volume} {183}},\ \bibinfo {pages} {1269--1289} (\bibinfo {year} {2012}{\natexlab{b}})}\BibitemShut {NoStop}%
\bibitem [{\citenamefont {Qiu}\ \emph {et~al.}(2021{\natexlab{b}})\citenamefont {Qiu}, \citenamefont {Cohen}, \citenamefont {Novichkova},\ and\ \citenamefont {Refaely-Abramson}}]{qiu2021signatures}%
  \BibitemOpen
  \bibfield  {author} {\bibinfo {author} {\bibfnamefont {Diana~Y}\ \bibnamefont {Qiu}}, \bibinfo {author} {\bibfnamefont {Galit}\ \bibnamefont {Cohen}}, \bibinfo {author} {\bibfnamefont {Dana}\ \bibnamefont {Novichkova}}, \ and\ \bibinfo {author} {\bibfnamefont {Sivan}\ \bibnamefont {Refaely-Abramson}},\ }\bibfield  {title} {\enquote {\bibinfo {title} {Signatures of dimensionality and symmetry in exciton band structure: Consequences for exciton dynamics and transport},}\ }\href@noop {} {\bibfield  {journal} {\bibinfo  {journal} {Nano letters}\ }\textbf {\bibinfo {volume} {21}},\ \bibinfo {pages} {7644--7650} (\bibinfo {year} {2021}{\natexlab{b}})}\BibitemShut {NoStop}%
\bibitem [{\citenamefont {Ambegaokar}\ and\ \citenamefont {Kohn}(1960)}]{ambegaokar1960electromagnetic}%
  \BibitemOpen
  \bibfield  {author} {\bibinfo {author} {\bibfnamefont {Vinay}\ \bibnamefont {Ambegaokar}}\ and\ \bibinfo {author} {\bibfnamefont {Walter}\ \bibnamefont {Kohn}},\ }\bibfield  {title} {\enquote {\bibinfo {title} {Electromagnetic properties of insulators. i},}\ }\href@noop {} {\bibfield  {journal} {\bibinfo  {journal} {Physical Review}\ }\textbf {\bibinfo {volume} {117}},\ \bibinfo {pages} {423} (\bibinfo {year} {1960})}\BibitemShut {NoStop}%
\bibitem [{\citenamefont {Gatti}\ and\ \citenamefont {Sottile}(2013)}]{gatti2013exciton}%
  \BibitemOpen
  \bibfield  {author} {\bibinfo {author} {\bibfnamefont {Matteo}\ \bibnamefont {Gatti}}\ and\ \bibinfo {author} {\bibfnamefont {Francesco}\ \bibnamefont {Sottile}},\ }\bibfield  {title} {\enquote {\bibinfo {title} {Exciton dispersion from first principles},}\ }\href@noop {} {\bibfield  {journal} {\bibinfo  {journal} {Physical Review B—Condensed Matter and Materials Physics}\ }\textbf {\bibinfo {volume} {88}},\ \bibinfo {pages} {155113} (\bibinfo {year} {2013})}\BibitemShut {NoStop}%
\bibitem [{\citenamefont {Egerton}(2008)}]{egerton2008electron}%
  \BibitemOpen
  \bibfield  {author} {\bibinfo {author} {\bibfnamefont {Ray~F}\ \bibnamefont {Egerton}},\ }\bibfield  {title} {\enquote {\bibinfo {title} {Electron energy-loss spectroscopy in the tem},}\ }\href@noop {} {\bibfield  {journal} {\bibinfo  {journal} {Reports on Progress in Physics}\ }\textbf {\bibinfo {volume} {72}},\ \bibinfo {pages} {016502} (\bibinfo {year} {2008})}\BibitemShut {NoStop}%
\bibitem [{\citenamefont {Kuzmany}(2009)}]{kuzmany2009solid}%
  \BibitemOpen
  \bibfield  {author} {\bibinfo {author} {\bibfnamefont {Hans}\ \bibnamefont {Kuzmany}},\ }\href@noop {} {\emph {\bibinfo {title} {Solid-state spectroscopy: an introduction}}}\ (\bibinfo  {publisher} {Springer Science \& Business Media},\ \bibinfo {year} {2009})\BibitemShut {NoStop}%
\end{thebibliography}%

\end{document}